\documentclass[journal]{IEEEtran}
\usepackage{graphicx}
\usepackage{amsfonts}
\usepackage{amsmath} 
\usepackage{amssymb} 
\usepackage{algorithmic}
\usepackage{array}
\usepackage{cite}
\usepackage{cases}
\usepackage{epstopdf}
\usepackage{enumerate}
\usepackage{lipsum}
\usepackage[pagewise]{lineno}
\usepackage{multicol}
\usepackage{paralist}
\usepackage{subfigure}  
\usepackage{subeqnarray}
\usepackage{stfloats}
\usepackage{setspace}
\usepackage{textcomp}
\usepackage{url,epsfig,bm,dsfont}
\usepackage{hyperref}

\begin{document}

\title{Designing and Analysis of A Wi-Fi Data Offloading Strategy Catering for the Preference of Mobile Users}

\author{\IEEEauthorblockN{Xiaoyi Zhou, Tong Ye, \IEEEmembership{Member,~IEEE,} and Tony T. Lee, \IEEEmembership{Fellow,~IEEE}}}

\maketitle

\begin{abstract}
In recent years, offloading mobile traffic through Wi-Fi has emerged as a potential solution to lower down the communication cost for mobile users. Users hope to reduce the cost while keeping the delay in an acceptable range through Wi-Fi offloading. Also, different users have different sensitivities to the cost and the delay performance. How to make a proper cost-delay tradeoff according to the user's preference is the key issue in the design of the offloading strategy. To address this issue, we propose a preference-oriented offloading strategy for current commercial terminals, which transmit traffic only via one channel simultaneously. We model the strategy as a three-state M/MMSP/1 queueing system, of which the service process is a Markov modulated service process (MMSP), and obtain the structured solutions by establishing a hybrid embedded Markov chain. Our analysis shows that, given the user's preference, there exists an optimal deadline to maximize the utility, which is defined as the linear combination of the cost and the delay. We also provide a method to select the optimal deadline. Our simulation demonstrates that this strategy with the optimal deadline can achieve a good performance.
\end{abstract}


\section{Introduction}
In recent years, the explosion of mobile traffic makes the cellular network overloaded, which degrades users' satisfaction. One way to expand the capacity of the cellular network is to install more and more base stations. However, owing to the scarcity of the licensed frequency band (LFB) \cite{04} and the high cost of base stations, the growth rate of the cellular capacity can hardly catch up with the bandwidth demand of mobile data.

Meanwhile, Wi-Fi (Wireless Fidelity) technology sheds some light on the expansion of wireless capacity. Its installation cost is low and the construction time is short. More importantly, Wi-Fi can work on the free unlicensed frequency band (UFB) \cite{05}. Offloading mobile traffic through Wi-Fi has been a widely accepted solution to achieve low communication costs for both operators and users \cite{02}. Nowadays, more and more Wi-Fi hotspots have been installed in public areas to provide cheap bandwidth for mobile users. With the increase of Wi-Fi hotspots, an open and shared Wi-Fi wireless environment is coming true \cite{19}.

Currently, mobile users moving in the city pass through the coverages of cellular networks and Wi-Fi hotspots alternatively. In response to such a wireless environment, most mobile terminals offload data in the following way by default \cite{07}: the terminal transmits data via the Wi-Fi when there is a Wi-Fi hotspot available, and sends data via the cellular network once the Wi-Fi is lost. Such an offloading strategy is known as on-the-spot offloading \cite{22,13}. Though this strategy has a small communication delay, it leads to a low offloading efficiency, which is defined as the ratio of the mobile data offloaded via the Wi-Fi. A low offloading efficiency signifies a high communication cost.

A straightforward strategy to maximize the offloading efficiency is to pause data transmission once the terminal loses the Wi-Fi, and resume data transmission when the Wi-Fi is available. This strategy is referred to as pure offloading in this paper. Clearly, this strategy can achieve high offloading efficiency, but it will incur a large delay.

However, users desire the strategy that can make a proper compromise between the delay performance and the offloading efficiency according to their requirements. In this case, either the on-the-spot offloading strategy or the pure offloading strategy may not be a good choice. To address this issue, \cite{01} proposed an opportunistic Wi-Fi data offloading strategy based on a utility function, which was defined as the combination of delay and cost. Herein, weights are multiplied with delay and cost to model the user's sensitivity to them. In this strategy, each time the data generate or the terminal loses the Wi-Fi during data transmission, it estimates the utility of the action that it immediately switches to the cellular network and the action that it waits for the next Wi-Fi. This strategy always selects the one with larger estimated utility and thus can achieve a higher utility than the on-the-spot offloading and the pure offloading in most cases. However, this strategy requires real-time estimation according to the prior and the real-time information about the interval time between Wi-Fi connections. Real-time information collection and calculation will consume extra energy, thus may not be suitable for energy-constrained terminals.

Even so, the opportunistic offloading strategy in \cite{01} provides a good idea for the performance tradeoff. The essential of it is to adaptively implement the on-the-spot offloading and the pure offloading such that they appear with a certain probability. From the statistical perspective, the data transmission is deferred for a while on average when decisions have to be made. This hints that, if a proper deferred time can be found, it is possible to obtain a high utility without real-time monitoring and calculation.
\subsection{Our approach and contributions}
We propose a preference-oriented Wi-Fi offloading strategy for current commercial mobile terminals in the paper. Our goal is to achieve a high utility in the long run while avoiding real-time monitoring and calculation. In our strategy, there is a preset deadline for the transmission server. Once the terminal loses Wi-Fi, it pauses data transmission. If the terminal connects to a new Wi-Fi hotspot before the deadline expires, it will resume data transmission via Wi-Fi; otherwise, it will transmit the data through the cellular network when the deadline expires.

We model the proposed strategy as an M/MMSP/1 queueing system with three service states. To circumvent the dependency among the service times in this model, we establish a hybrid embedded Markov chain, in which both the epoch when a data frame begins its service and that when the service state transits to another are considered as embedded points. We derive the probability that a frame starts its service in each service state and the mean service time, and finally solve the structured expression of the mean delay and the offloading efficiency.

Our analytical results indicate that in the proposed strategy, the offloading efficiency is improved at the expense of the delay performance. In particular, with the increase of deadline, the mean delay grows faster than the offloading efficiency. These properties imply that there exists an optimal deadline that can trade the delay performance for the offloading efficiency such that the utility is maximized. Accordingly, we provide the method to seek the optimal deadline. We compare our strategy with the on-the-spot offloading and the pure offloading through simulations, which show that our strategy can achieve a higher utility than these two extreme strategies.

The rest of this paper is organized as follows. In Section \ref{percedure}, we propose a Wi-Fi offloading strategy for current commercial terminals and model the strategy as a three-state M/MMSP/1 queueing system. In Section \ref{embedded}, we develop a hybrid embedded Markov chain to analyze this model. In Section \ref{properties}, we solve the mean delay and the offloading efficiency, and observe the system properties. Section \ref{applications} shows how to seek the optimal deadline. Also, we demonstrate that our strategy with the optimal deadline can achieve a higher utility than the on-the-spot offloading and the pure offloading. We discuss some related works in Section \ref{related}. Section \ref{conclusion} concludes this paper.
\section{Preference-oriented offloading procedure}\label{percedure}
\begin{figure}[htp]
\centering
\includegraphics[scale=0.6]{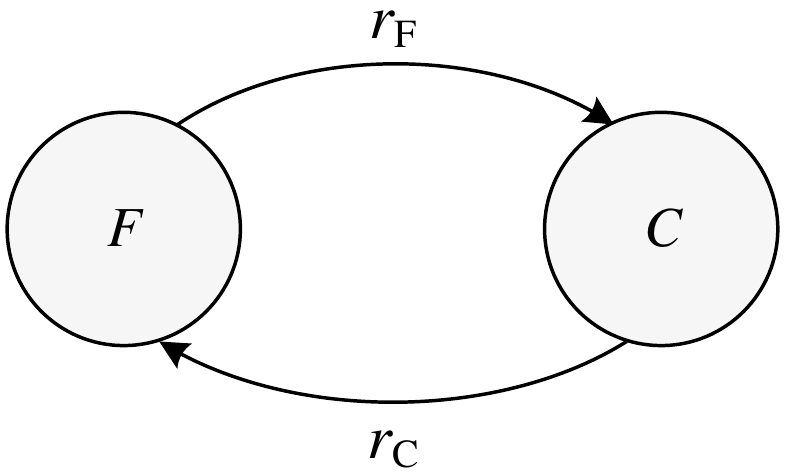}
\caption{Transition of wireless channel states in urban areas.}
\label{fig1}
\end{figure}

Nowadays, the cellular network is nearly ubiquitous in the city, while the Wi-Fi hotspots are distributed with numerous small areas. When a terminal is moving in urban areas, it alternately passes through the Wi-Fi coverage and the cellular coverage. Thus, the terminal perceives the wireless channel periodically switching between two states, as Figure \ref{fig1} illustrates, where $C$ denotes the channel state that there is the cellular signal only and $F$ denotes the state that the Wi-Fi signal is available. 

Define a $C$ state followed by an $F$ state as a channel cycle. Let $1/r_F$ and $1/r_C$ be the average duration times of the $F$ state and the $C$ state, respectively. The average channel-cycle time is $1/r_F+1/r_C$. It follows that $r_F$ is the transition rate from state $F$ to state $C$, and $r_C$ is that from state $C$ to state $F$.

Once the deployment of Wi-Fi hotspots is given, the ratio of $1/r_F$ and $1/r_C$ is fixed. In this case, \cite{09} defines $R=(1/r_F)/(1/r_C+1/r_F )=r_C/(r_C+r_F)$ as the Wi-Fi availability ratio to characterize the time fraction that a terminal can access the Wi-Fi hotspots. If the density of Wi-Fi hotspots is high, the Wi-Fi availability ratio $R$ and the opportunity for data offloading are high.
\subsection{Offloading procedure}\label{2.1}
To reduce the communication cost, the current commercial mobile terminals automatically transmit data via Wi-Fi when there is a Wi-Fi hotspot available. According to this characteristic, we propose a preference-oriented Wi-Fi offloading strategy as follows.

When there is an available Wi-Fi hotspot, the terminal offloads the data via Wi-Fi. Once the terminal loses the Wi-Fi, it pauses the transmission to wait for the next Wi-Fi hotspot and sets a deadline, denoted by $\tau$, at the same time. If the terminal can set up a new Wi-Fi connection before the deadline expires, it resumes data offloading via Wi-Fi; otherwise, it starts the transmission via the cellular network once the deadline expires.

\begin{figure}[htp]
\centering
\includegraphics[scale=0.5]{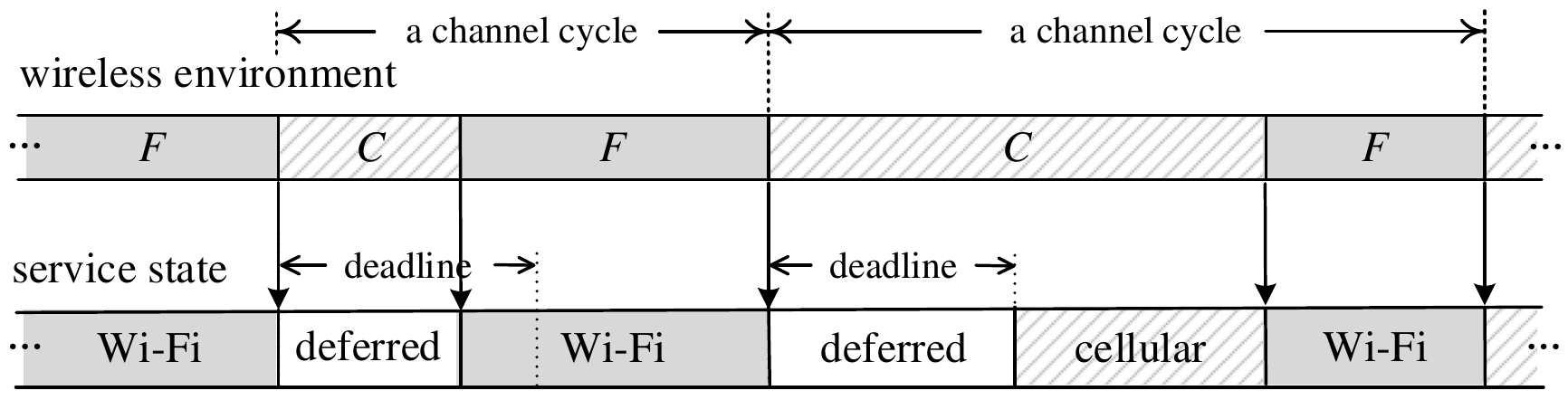}
\caption{Process of the proposed Wi-Fi offloading strategy.}
\label{fig2}
\end{figure}

The terminal with this strategy has three transmission (or service) states: (1) deferred state (or state 0), the transmission is paused and thus transmission rate is $\mu_0=0$, (2) cellular state (or state 1), transmission via the cellular network and the transmission rate is $\mu_1$ (frames/s), and (3) Wi-Fi state (or state 2), transmission via the Wi-Fi and the transmission rate is $\mu_2$ (frames/s). The transition of three service states is delineated in Figure \ref{fig2}.

Based on different values of deadline $\tau$, our strategy can reduce to the on-the-spot offloading or the pure offloading. When $\tau=0$, the terminal will transmit data via the cellular network once it loses the Wi-Fi signal, and our strategy changes to the on-the-spot offloading. When $\tau\rightarrow \infty$, our strategy will offload all the data via Wi-Fi and thus is equivalent to the pure offloading.
\subsection{Utility function}\label{2.2}
Different users have different sensitivities to the delay performance and the communication cost. The purpose of the offloading strategy is to make a tradeoff according to the user's sensitivity. We define the utility as the linear combination of the mean delay of frames and the offloading efficiency in this paper, since the data is transmitted in the form of frames in practice. Herein, the delay of a frame defines the duration time from the epoch a frame generates to the epoch it is completely transmitted. Our goal is to find a proper deadline for our strategy, such that it can meet the users' requirements on the delay and the cost. Let $D$ be the mean delay of the frames, $\hat{D}$ be the maximal mean delay for all the values of $\tau$, and $\eta$ be the offloading efficiency. We define the utility function as
\begin{equation}
\label{eq1}
U=1-a(D/\hat{D})-(1-a)(1-\eta),
\end{equation}
where $0<D/\hat{D}\le1$ is the normalized delay, $1-\eta$ stands for the communication cost, and $0\le a \le1$ is the preference weight. The value range of utility $U$ is $[0,1]$. The utility declines with the increase of the normalized mean delay or the cost, which is corresponding to the fact that the satisfaction of the user will be lowered down if the mean delay or the cost is high. The preference weight $a$ indicates the user's sensitivity to the delay and the cost. In practice, the value of $a$ can be specified by the user. If $a$ is large, the user is more sensitive to the delay and the utility decreases fast with the delay; otherwise, the user cares more about the cost.
\subsection{M/MMSP/1 queueing model}\label{2.3}
To facilitate the analysis, we make the following assumptions:
\begin{enumerate}[1)]
\item The duration times of channel states $C$ and $F$ are exponential random variables with mean $1/r_C$ and $1/r_F$ \cite{01,26};\label{a1}
\item The deadline is an exponential random variable with mean $\tau$;\label{a2}
\item The arrival process of frames is a Poisson process with rate $\lambda$ frames/s;\label{a3}
\item The frame size is exponentially distributed;\label{a4}
\item The buffer size is infinite and the frames are served in a first-in-first-out manner.\label{a5}
\end{enumerate}
\begin{figure}[htp]
\centering
\includegraphics[scale=0.7]{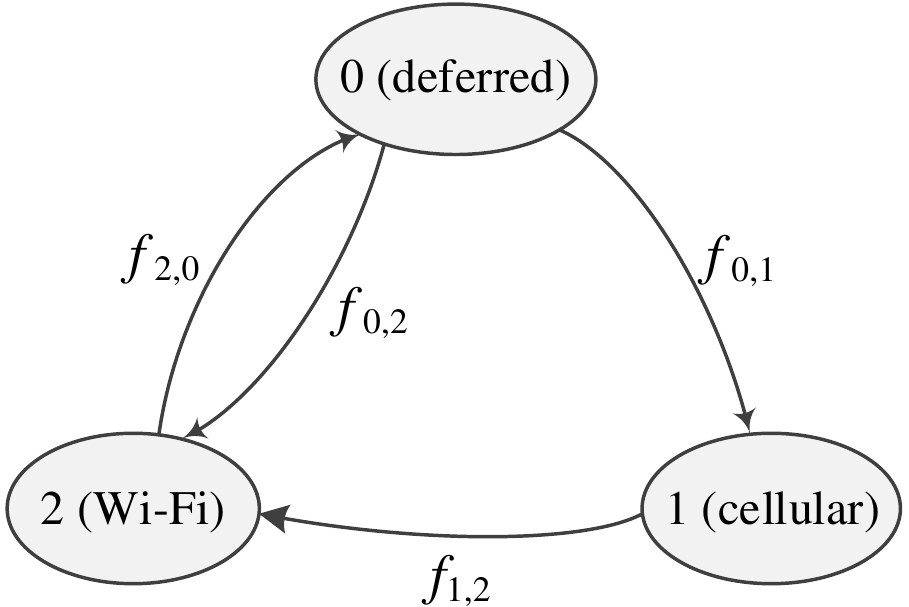}
\caption{State transition of the data transmission.}
\label{fig3}
\end{figure}

With assumptions \ref{a1}) and \ref{a2}), the data transmission process of the terminal can be considered as a Markov modulated service process (MMSP) \cite{15} with three service states in Figure \ref{fig3}. Let $f_{i,j}$ be the transition rate from state $i$ to state $j$, where $i,j=0,1,2$. According to Figure \ref{fig1} and \ref{fig2}, $f_{i,j}$ is given by
\begin{align}
&f_{0,2}=f_{1,2}=r_C,\\
&f_{2,0}=r_F,\\
&f_{0,1}=1/\tau.
\end{align}
It follows that the steady-state probabilities of three service states in Figure \ref{fig3} are given by
\begin{subequations}
\begin{align}
&\pi_0=(1-R)  \frac{r_C \tau}{r_C \tau+1},\\
&\pi_1=(1-R)  \frac{1}{r_C \tau+1},\\
&\pi_2=R.
\end{align}
\end{subequations}
Thus, the capacity that the terminal can offer is:
\begin{equation}\label{eq11}
\hat{\mu}=\pi_1 \mu_1+\pi_2 \mu_2=\frac{1-R}{r_C \tau+1} \mu_1+R\mu_2,
\end{equation}
which decreases with $\tau$. As \cite{25} and \cite{24} demonstrate, the mean service time of frames is lower bounded by $1/\hat{\mu}$. Recall that the probability that the system is busy is the product of the input traffic rate $\lambda$ and the mean service time. One can expect that, for a fixed $\lambda$, increasing $\tau$ will increase the mean service time, and thus the fraction of the time that the system is busy. This hints that a large $\tau$ will degrade the delay performance.

Combining assumptions \ref{a3}) through \ref{a5}), the offloading process can be modeled as an M/MMSP/1 queue with three service states. As Appendix \ref{appendixA} explains, such M/MMSP/1 queue can be analyzed by a two-dimension continuous-time Markov chain, from which we can numerically solve the mean delay. However, attributing to the complexity of the system with multiple service states, this solution is an intricate combination of mathematical variables, which supplies little help in exploring the system performance. Thus, we develop a new approach to solve this M/MMSP/1 queue in Section \ref{embedded}.
\section{Hybrid embedded Markov chain}\label{embedded}
It is well known that the analysis of the M/MMSP/1 queuing systems with multiple service states is very difficult, the difficulty of which mainly lies in the fact that the service time of a frame is related to the service state when its service starts \cite{25,28}. To address this issue, Section \ref{3.1} develops a hybrid embedded Markov chain, based on which Section \ref{3.2} derives the probability that a frame starts its service in a service state. Using such a probability, we derive the service time in Section \ref{3.3}.

\subsection{Embedded points}\label{3.1}
To delineate the dependencies among service times, we develop a hybrid embedded Markov chain to solve the system, in which two types of time points are embedded into the data offloading process. We consider the epoch when a frame starts its service, since the service time of frames depends on the service state at this epoch. We also observe the epoch at the transition of service states, since the dependency of the service time is essentially caused by the service state transitions during the service of a frame. We thus define two types of embedded points as follows:
\begin{enumerate}[1)]
\item State-transition point $\Phi_j$: epoch when the service state transits to state $j$;
\item Start-service point $S_j$: epoch when a frame starts its service and the service state is $j$,
\end{enumerate}
where $j=0,1,2$. Clearly, the time interval between two adjacent embedded points is exponentially distributed.

\begin{figure}[htp]
\centering
\subfigure[current state $j=0$ ]{
 \label{fig4a}
 \includegraphics[scale=0.8]{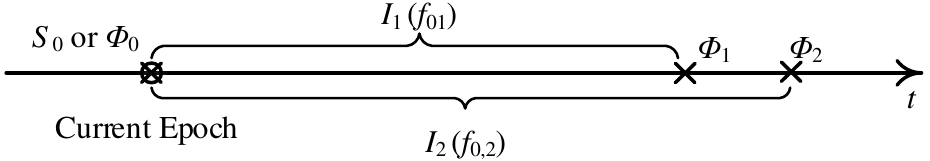}}
\subfigure[current state $j=1$]{{}
 \label{fig4b}
 \includegraphics[scale=0.8]{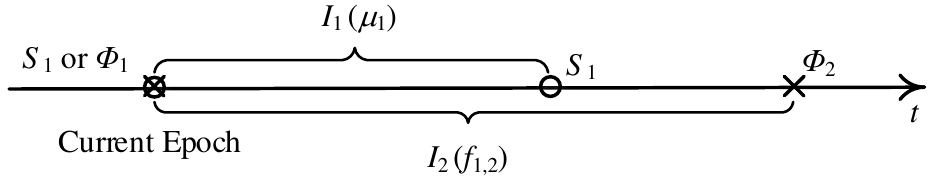}}
 \subfigure[current state $j=2$]{
 \label{fig4c}
 \includegraphics[scale=0.8]{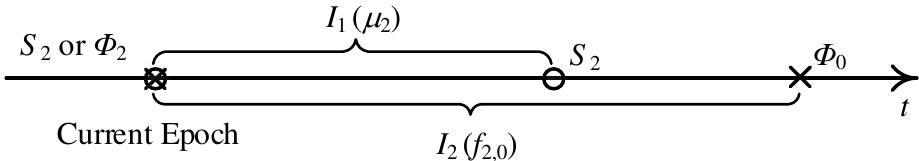}}
\caption{Relationship between two kinds of embedded points.}
\label{fig4}
\end{figure}

Suppose the current epoch is an embedded point of which the service state is the deferred state, i.e., $j=0$, as Figure \ref{fig4a} shows. Since the service is suspended at the current epoch, the next event may be a state transition from service state $0$ to service state $i$ after time $I_i$ which is an exponential random variable with parameter $f_{0,i}$, where $i=1,2$. Thus, the type of the next embedded point is determined by which kind of the service state transition happens first. It follows that the distribution of the time $I=\min\limits_{i}I_i$ from the current point to the next point is exponentially distributed with parameter $\sum_{i=1}^{2}f_{0,i}$, and the next embedded point is $\Phi_i$ with probability $f_{0,i}/\sum_{i=1}^{2}f_{0,i}$, where $i=1,2$.

Similarly, when the current epoch is an embedded point of which the service state is state $j>0$, the next embedded point will be $\Phi_{\overline{j}}$ with probability $f_{j,\overline{j}}/\big(f_{j,\overline{j}}+\mu_{j}\big)$ or $S_j$ with probability $\mu_{j}/\big(f_{j,\overline{j}}+\mu_{j}\big)$, where $\overline{j}\triangleq\left(j+1\right)mod$ 3, as Figure \ref{fig4b} and \ref{fig4c} show. Also, the distribution of the time interval from the current point to the next point is exponentially distributed with parameter $f_{j,\overline{j}}+\mu_{j}$.
\subsection{Start service probability}\label{3.2}
\begin{figure}[htp]
\centering
\subfigure[]{
 \label{fig5a}
 \includegraphics[scale=1.3]{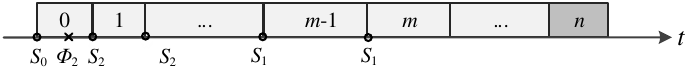}}
\subfigure[]{{}
 \label{fig5b}
 \includegraphics[scale=1.3]{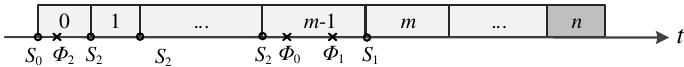}}
\caption{The $m$th frame starts service in the cellular state, while the last event is (a) the $(m-1)$th frame starts its service in the cellular state, or (b) the service state transits to cellular state when the $(m-1)$th frame is in service.}
\label{fig5}
\end{figure}

The start service probability $\hat{\Phi}_j$ is defined as the probability that a data frame starts its service in state $j$. Consider a newly-arrived frame, which sees $n$ frames in the buffer. These frames are labeled according to their sequence in the queue. The head-of-line (HOL) frame is labeled with 0 and the newly arrived-frame is labeled with $n$, as Figure \ref{fig5} shows. We define two types of conditional probabilities corresponding to the embedded points:
\begin{enumerate}[1)]
\item $\hat{\pi}_{n,j}\!\left(m\right)=Pr$\{the $m$th data frame starts its service in service state $j$ $|$ the new arrival sees $n$ frames in the buffer\}
\item $\hat{\varphi}_{n,j}\!\left(m\right)=Pr$\{the service state transits to state $j$ when the $m$th frame is in service $|$ the new arrival sees $n$ frames in the buffer\}
\end{enumerate}
where $m=0,1,2,\cdots,n$.

$\hat{\pi}_{n,j}\!\left(m\right)$ is defined on start-service point $S_j$, at which the $(m-1)$th frame finishes its service when the service state is $j$, for $m=1,2,\cdots,n$. When the service state is the deferred state, i.e., $j=0$, the $(m-1)$th frame cannot finish the service, since the service rate is 0. Therefore, it is impossible that the $m$th frame starts the service in the deferred state, i.e.,
\begin{subequations}\label{eq12}
\begin{gather}
\hat{\pi}_{n,0}\!\left(m\right)=0.\\
\intertext{\indent In other words, the service of the $m$th frame can only start in the cellular state ($j=1$) or the Wi-Fi state ($j=2$). Consider the case that the $m$th frame starts its service in the cellular state. The last event may be either that the ($m-1$)th frame starts its service in the cellular state, as Figure \ref{fig5a} shows, or that the service state transits to the cellular state when the ($m-1$)th packet is in service, as Figure \ref{fig5b} plots. Recall that the probability that the next embedded point is $S_1$ on which a new service starts in the cellular state is $\mu_1/\left(\mu_1+f_{1,2}\right)$, given that the current service state is the cellular state ($j=1$). It follows that the probability $\hat{\pi}_{n,1}\!\left(m\right)$ can be given by}
\hat{\pi}_{n,1}\!\left(m\right)=\frac{\mu_1}{\mu_{1}\!+\!f_{1,2}}\Big[\hat{\pi}_{n,1}\!\left(m\!-\!1\right)\!+\!\hat{\varphi}_{n,1}\!\left(m\!-\!1\right)\Big].\\
\intertext{Similarly, we can obtain the probability $\hat{\pi}_{n,2}\!\left(m\right)$ by}
\hat{\pi}_{n,2}\!\left(m\right)=\frac{\mu_2}{\mu_2\!+\!f_{2,0}}\Big[\hat{\pi}_{n,2}\!\left(m\!-\!1\right)\!+\!\hat{\varphi}_{n,2}\!\left(m\!-\!1\right)\Big].
\end{gather}
\end{subequations}

$\hat{\varphi}_{n,j}\!\left(m\right)$ is defined on state-transition point $\Phi_j$. Since the service state changes after the state-transition point, the last embedded point happens when the $m$th frame is being served in other states. Consider the embedded point when the $m$th frame is in service and the service state transits to the deferred state, i.e., $j=0$. Since the deferred state can only be accessed by the Wi-Fi state ($j=2$), the last event can be either the $m$th frame starts its service in the Wi-Fi state, or the service state transits to the Wi-Fi state when the $m$th frame is in service. Recall that the probability that the next embedded point is $\Phi_0$ on which the service state transits to the deferred state is $f_{2,0}/(\mu_2+f_{2,0})$, given that the current service state is the Wi-Fi state. It follows that the probability $\hat{\varphi}_{n,0}\!\left(m\right)$ can be written as
\begin{subequations}\label{eq13}
\begin{gather}
\hat{\varphi}_{n,0}\!\left(m\right)=\frac{f_{2,0}}{\mu_2+f_{2,0}}\Big[\hat{\pi}_{n,2}\!\left(m\right)\!+\!\hat{\varphi}_{n,2}\!\left(m\right)\Big].
\end{gather}
Similarly, $\hat{\varphi}_{n,1}\!\left(m\right)$ and $\hat{\varphi}_{n,2}\!\left(m\right)$ can be given by
\begin{align}
\hat{\varphi}_{n,1}\!\left(m\right)=&\frac{f_{0,1}}{f_{0,1}\!+\!f_{0,2}}\Big[\hat{\pi}_{n,0}\!\left(m\right)\!+\!\hat{\varphi}_{n,0}\!\left(m\right)\Big],\\
\begin{split}
\hat{\varphi}_{n,2}\!\left(m\right)=&\frac{f_{0,2}}{f_{0,1}\!+\!f_{0,2}}\Big[\!\hat{\pi}_{n,0}\!\left(m\right)\!+\!\hat{\varphi}_{n,0}\!\left(m\right)\Big]\!\\
&+\!\frac{f_{1,2}}{\mu_1\!+\!f_{1,2}}\Big[\!\hat{\pi}_{n,1}\!\left(m\right)\!+\!\hat{\varphi}_{n,1}\!\left(m\right)\Big]\!.
\end{split}
\end{align}
\end{subequations}

Combing (\ref{eq12}) and (\ref{eq13}), we have the relations between $\hat{\pi}_{n,j}\!\left(m\right)$ and $\hat{\pi}_{n,j}\!\left(m-1\right)$:
\begin{equation}\label{eq14}
\begin{pmatrix}
  \hat{\pi}_{n,0}\!\left(m\right) \\
  \hat{\pi}_{n,1}\!\left(m\right) \\
  \hat{\pi}_{n,2}\!\left(m\right)
\end{pmatrix}
=\hat{Q}
\begin{pmatrix}
  \hat{\pi}_{n,0}\!\left(m-1\right) \\
  \hat{\pi}_{n,1}\!\left(m-1\right) \\
  \hat{\pi}_{n,2}\!\left(m-1\right)
\end{pmatrix}
=\hat{Q}^{m}
\begin{pmatrix}
  \hat{\pi}_{n,0}\!\left(0\right) \\
  \hat{\pi}_{n,1}\!\left(0\right) \\
  \hat{\pi}_{n,2}\!\left(0\right)
\end{pmatrix},
\end{equation}
where the coefficient matrix $\hat{Q}$ is
\begin{equation}\label{eq15}
\begin{split}
&\hat{Q}=\\
&\begin{pmatrix}
 \begin{smallmatrix}
  0 & 0 & 0 \\
  \beta \frac{f_{0,1}}{f_{0,1}\!+\!f_{0,2}}\!\left(1\!+\!\frac{f_{2,0}}{\mu_2}\right) & \beta \left(1\!+\!\frac{f_{0,1}}{f_{0,1}\!+\!f_{0,2}}\frac{f_{2,0}}{\mu_2}\right) & \beta\frac{f_{0,1}}{f_{0,1}\!+\!f_{0,2}}\!\frac{f_{2,0}}{\mu_2}\\
  \beta\left(\frac{f_{1,2}}{\mu_1}+\frac{f_{0,2}}{f_{0,1}\!+\!f_{0,2}}\right) & \beta\frac{f_{1,2}}{\mu_1} & \beta \left(1+\frac{f_{1,2}}{\mu_1}\right)
\end{smallmatrix}
 \end{pmatrix}
\end{split}
\end{equation}
and
\begin{equation}\label{eq16}
\beta=\frac{\left(r_C\tau+1\right)\mu_1\mu_2}{r_F\mu_1+\left(r_C\tau+1\right)r_C\mu_2+\left(r_C\tau+1\right)\mu_1\mu_2}.
\end{equation}
Solving (\ref{eq14}), we obtain
\begin{subequations}\label{eq17}
\begin{align}
&\hat{\pi}_{n,0}\!\left(m\right)=0\\
\begin{split}
\hat{\pi}_{n,1}\!\left(m\right)=\theta_1+\Big(\theta_2-\frac{f_{0,2}}{f_{0,1}+f_{0,2}}\Big)\beta^{m}\hat{\pi}_{n,0}\!\left(0\right)\!\\
+\theta_2\beta^{m}\hat{\pi}_{n,1}\!\left(0\right)-\theta_1\beta^{m}\hat{\pi}_{n,2}\!\left(0\right)
\end{split}\\
\begin{split}
\hat{\pi}_{n,2}\!\left(m\right)=\theta_2-\Big(\theta_2-\frac{f_{0,2}}{f_{0,1}+f_{0,2}}\Big)\beta^{m}\hat{\pi}_{n,0}\!\left(0\right)\!\\
-\theta_2\beta^{m}\hat{\pi}_{n,1}\!\left(0\right)+\theta_1\beta^{m}\hat{\pi}_{n,2}\!\left(0\right),
\end{split}
\end{align}
\end{subequations}
where $m=1,2,\cdots,n$ and
\begin{equation}\label{eq18}
\theta_{j}=\frac{\pi_j\mu_j}{\sum_{j=1}^{2}\pi_j\mu_j}
\end{equation}
is the ratio of the capacity that service state $j$ can provide, where $j=1,2$.

As for $m=0$, $\hat{\pi}_{n,j}\!\left(0\right)$ is the probability that the HOL frame starts its service when the service state is $j$, given that the newly-arrived frame sees $n$ frames in the buffer. According to the PASTA property \cite{27}, $\hat{\pi}_{n,j}\!\left(0\right)=p_{n,j}/p_n$, where $p_{n,j}$ is the steady-state probability that there are $n$ frames in the buffer and the service state is $j$, and $p_n=\sum_{j=0}^{2}p_{n,j}$ is the steady-state probability that there are $n$ frames in the buffer.

By definition, a newly-arrived frame that sees $n$ frames in the buffer upon its arrival starts its service in state $j$ is $\hat{\pi}_{n,j}\left(n\right)$. Thus, the start service probability $\hat{\pi}_j$ is
\begin{equation}\label{eq19}
  \hat{\pi}_j=\sum_{n=0}^{\infty}p_n\hat{\pi}_{n,j}\left(n\right).
\end{equation}

Combining (\ref{eq17}) and (\ref{eq19}), we have the following results:
\begin{subequations}\label{eq20}
\begin{align}
\hat{\pi}_0=p_{0,0}\\
\begin{split}
\hat{\pi}_1=\theta_1+&\Big(\theta_2-\frac{r_C\tau}{r_C\tau+1}\Big)G_0\left(\beta\right)\\
&+\theta_2G_1\left(\beta\right)-\theta_1G_2\left(\beta\right)\!-\!\frac{1}{r_C\tau+1}p_{0,0}
\end{split}\\
\begin{split}
\hat{\pi}_2=\theta_2-&\Big(\theta_2-\frac{r_C\tau}{r_C\tau+1}\Big)G_0\left(\beta\right)\\
&-\theta_2G_1\left(\beta\right)+\theta_1G_2\left(\beta\right)\!-\!\frac{1}{r_C\tau+1}p_{0,0},
\end{split}
\end{align}
\end{subequations}
where $G_j\!\left(z\right)\!=\!\sum_{n=0}^{\infty}p_{n,j}z^n$ and $p_{0,0}$ can be derived by the two-dimensional continuous-time Markov chain presented in Appendix \ref{appendixA}.
\subsection{Mean service time}\label{3.3}
\begin{figure}[htp]
\centering
\includegraphics[scale=0.48]{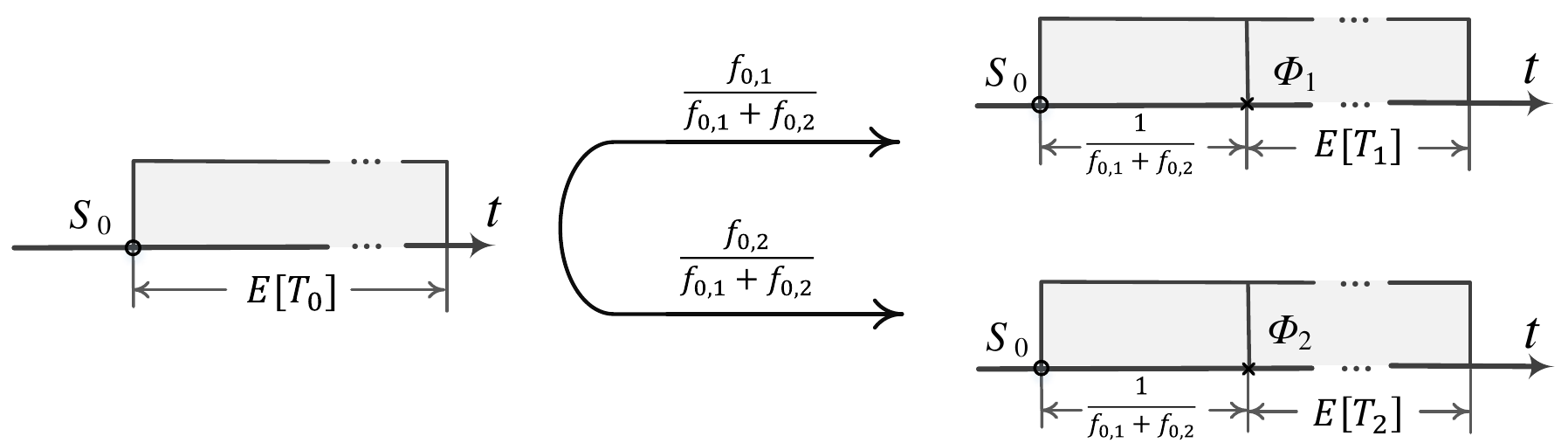}
\caption{The service time when a frame starts its service in the deferred state.}
\label{fig6}
\end{figure}
Let $T_j$ be the service time needed to serve a frame if the frame starts its service in service state $j$ \cite{29}. Consider a frame that the system is empty and in the deferred state ($j=0$) upon its arrival. This epoch corresponds to an embedded point $S_0$, which starts a service time $T_0$ as shown in Figure \ref{fig6}. In this state, the terminal does not transmit the frame. Thus, this service state will transit to either the cellular state ($j=1$) or the Wi-Fi state ($j=2$) in the next embedded point with probability $f_{0,i}/\sum_{i=1}^{2}f_{0,i}$, where $i=1,2$. After that, the time still needed to finish the service is $T_i$. Recall that the average time from the current embedded point to the next embedded point is $1/\sum_{i=1}^{2}f_{0,i}$. Thus, $E\left[T_0\right]$ can be given by:
\begin{subequations}\label{eq21}
\begin{equation}\label{eq21a}
\begin{split}
E\left[T_0\right]=&\frac{f_{0,1}}{f_{0,1}\!+\!f_{0,2}}\!\left(\frac{1}{f_{0,1}\!+\!f_{0,2}}\!+\!E\left[T_1\right]\!\right)\\
&+\!\frac{f_{0,2}}{f_{0,1}\!+\!f_{0,2}}\!\left(\frac{1}{f_{0,1}\!+\!f_{0,2}}\!+\!E\left[T_2\right]\!\right)\!.
\end{split}
\end{equation}

Similarly, we obtain $E\left[T_1\right]$ in (\ref{eq21b}) and $E\left[T_2\right]$ in (\ref{eq21c}).
\begin{align}
  &E\left[T_1\right]\!=\!\frac{\mu_1}{\mu_1\!+\!f\!_{1,2}}\frac{1}{\mu_1\!+\!f\!_{1,2}}\!+\!\frac{f\!_{1,2}}{\mu_1\!+\!f\!_{1,2}}\!\left(\!\frac{1}{\mu_1\!+\!f\!_{1,2}}\!+\!E\left[T_2\right]\!\right)\!\label{eq21b}\\
  &E\left[T_2\right]\!=\!\frac{\mu_2}{\mu_2\!+\!f\!_{2,0}}\frac{1}{\mu_2\!+\!f\!_{2,0}}\!+\!\frac{f\!_{2,0}}{\mu_2\!+\!f\!_{2,0}}\!\left(\!\frac{1}{\mu_2\!+\!f\!_{2,0}}\!+\!E\left[T_0\right]\!\right)\!\label{eq21c}
\end{align}
\end{subequations}

Solving (\ref{eq21}), we can derive $E\left[T_j\right]$ as follows
\begin{subequations}
\begin{align}
&E\left[T_0\right]\!=\!\frac{(r_C\!+\!r_F\!+\!\mu_2)(r_C \tau\!+\!\mu_1 \tau\!+\!1)}{r_F \mu_1\!+\!(r_C \tau\!+\!1)r_C \mu_2\!+\!(r_C \tau\!+\!1)\mu_1 \mu_2}\\
&E\left[T_1\right]\!=\!\frac{(r_C\!+\!r_F\!+\!\mu_2)(r_C \tau\!+\!1)}{r_F \mu_1\!+\!(r_C \tau\!+\!1)r_C \mu_2\!+\!(r_C \tau\!+\!1)\mu_1 \mu_2}\\
&E\left[T_2\right]\!=\!\frac{(r_C\!+\!r_F)(r_C \tau\!+\!1)\!+\!(1\!+\!r_C \tau\!+\!r_F \tau)\mu_1}{r_F \mu_1\!+\!(r_C \tau\!+\!1)r_C \mu_2\!+\!(r_C \tau\!+\!1)\mu_1 \mu_2},
\end{align}
\end{subequations}
and thus the mean service time:
\begin{equation}\label{eq23}
E\left[T\right]=\sum_{j=0}^{2}\hat{\pi}_jE\left[T_j\right].
\end{equation}
\section{Properties of the preference-oriented strategy}\label{properties}
The mean delay and the offloading efficiency are two criteria of the preference-oriented offloading strategy. Using the hybrid embedded Markov chain developed in Section \ref{embedded}, we drive the structured expression of the mean delay and the offloading efficiency in Section \ref{4.1}, and observe the properties of the proposed strategy in Section \ref{4.2}.
\subsection{Mean delay and offloading efficiency}\label{4.1}
Typically, the waiting time of a frame is calculated as the sum of the residual service time of the HOL frame and the service time of the frames waiting in the buffer when this frame arrives at the system \cite{23}. In the following, we show that the waiting time can be easily derived using the memoryless property of the developed hybrid embedded Markov chain.

We consider the newly-arrived frame in Figure \ref{fig5}. This new frame sees $n$ frames waiting in the buffer when it arrives. As time goes on, the terminal transmits the previous frames one by one. The position of this new frame moves forward in the buffer and finally becomes the HOL frame. This process accompanies with service completions and service state transitions, which affect the service time of the previous frames, as we show in Section \ref{embedded}. It follows that the waiting time of this new frame correlates to not only the number of previous frames waiting before it in the buffer, but also the service state transitions that it experiences. To derive the mean waiting time, we thus need to define the conditional waiting time that is associated with the position of the new frame in the buffer.

$W_{n,j}\!\left(k\right)$: The conditional expected time from the epoch when a newly-arrived frame becomes the $k$th frame ($k=0,1,\cdots,n$) in the queue while the service state is $j$ to the epoch when it becomes the HOL frame, given that it sees $n$ frames in the buffer when it arrives.

\begin{figure}[htp]
\centering
\subfigure[A new frame arrives]{
 \label{fig7a}
 \includegraphics[scale=1.3]{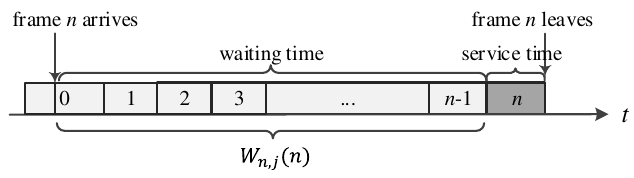}}
\subfigure[The newly-arrived frame becomes the $k$th in the buffer]{{}
 \label{fig7b}
 \includegraphics[scale=1.3]{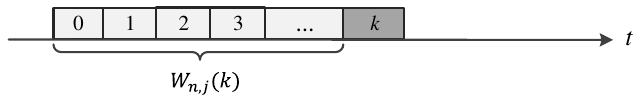}}
\caption{Waiting time of the newly-arrived frame.}
\label{fig7}
\end{figure}

By definition, $W_{n,j}\!\left(n\right)$ is the waiting time of the new frame given that it sees $n$ frames in the buffer and the service state is $j$ when it arrives, and $W_{n,j}\!\left(0\right)$ is zero. Figure \ref{fig7a} and \ref{fig7b} illustrate $W_{n,j}\!\left(n\right)$ and $W_{n,j}\!\left(k\right)$.

$W_{n,j}\!\left(k\right)$ starts at the epoch when the newly-arrived frame becomes the $k$th frame while the service state is $j$. If the next embedded point is a start-service point $S_j$, the HOL frame will finish its service in state $j$. In this case, the $k$th frame will be the ($k-1$)th frame in the queue and thus the expected remaining waiting time will be $W_{n,j}\!\left(k-1\right)$. If the next embedded point is a state-transition point $\Phi_i$, the service state will transit to state $i$ before the HOL frame finishes its service. In this case, the expected remaining waiting time will be $W_{n,i}\!\left(k\right)$, according to the memoryless property of the hybrid embedded Markov chain. Therefore, for $k=1,2,\cdots,n$, we have
\begin{subequations}
\begin{align}
\begin{split}
W_{n,0}\left(k\right)=&\frac{f_{0,1}}{f_{0,1}+f_{0,2}} \left[\frac{1}{f_{0,1}+f_{0,2}}+W_{n,1}\left(k\right)\right]\\
&+\frac{f_{0,2}}{f_{0,1}+f_{0,2}} \left[\frac{1}{f_{0,1}+f_{0,2}}+W_{n,2}\left(k\right)\right]
\end{split}\\
\begin{split}
W_{n,1}\left(k\right)=&\frac{\mu_1}{\mu_1+f_{1,2}}\left[\frac{1}{\mu_1+f_{1,2}}+W_{n,1}\left(k-1\right)\right]\\
&+\frac{f_{1,2}}{\mu_1+f_{1,2}} \left[\frac{1}{\mu_1+f_{1,2}}+W_{n,2}\left(k\right)\right]
\end{split}\\
\begin{split}
W_{n,2}\left(k\right)=&\frac{\mu_2}{\mu_2+f_{2,0}}\left[\frac{1}{\mu_2+f_{2,0}}+W_{n,2}\left(k-1\right)\right]\\
&+\frac{f_{2,0}}{\mu_2+f_{2,0}} \left[\frac{1}{\mu_2+f_{2,0}}+W_{n,0}\left(k\right)\right].
\end{split}
\end{align}
\end{subequations}

Rearranging above iteration equations into the matrix form, we have
\begin{equation}\label{eq25}
\begin{split}
\begin{pmatrix}
\!W_{n,0}\!\left(k\right)\! \\
  W_{n,1}\!\left(k\right)\! \\
  W_{n,2}\!\left(k\right)\!
\end{pmatrix}
=&\hat{Q}^{T}\!
\begin{pmatrix}
\!  W_{n,0}\!\left(k\!-\!1\right)\! \\
  W_{n,1}\!\left(k\!-\!1\right)\! \\
  W_{n,2}\!\left(k\!-\!1\right)\!
\end{pmatrix}
\!\!+\!\!
\begin{pmatrix}\!
  E\!\left[T_0\right]\! \\
  E\!\left[T_1\right]\! \\
  E\!\left[T_2\right]\!
\end{pmatrix}\\
=&\sum_{i=1}^{k-1}\!\left(\!\hat{Q}^{T}\!\right)\!^{\!i\!}\!
\begin{pmatrix}
\!  E\left[T_0\right]\! \\
  E\left[T_1\right]\! \\
  E\left[T_2\right]\!
\end{pmatrix}
\!\!+\!\!
\begin{pmatrix}
\!  E\left[T_0\right]\! \\
  E\left[T_1\right]\! \\
  E\left[T_2\right]\!
\end{pmatrix}\!\!,
\end{split}
\end{equation}
where
\begin{equation}\label{eq26}
\begin{split}
&\sum_{i=1}^{k-1}\!\left(\!\hat{Q}^{T}\!\right)\!^{\!i\!}=\\
&\begin{pmatrix}
\begin{smallmatrix}
0 & \theta_1\!\left(\!k\!-\!1\!\right)\!+\!\left(\!\theta_2\!-\!\frac{f_{0,2}}{f_{0,1}\!+\!f_{0,2}}\!\right)\!\frac{\beta\!-\!\beta^k}{1\!-\!\beta} & \theta_2\!\left(\!k\!-\!1\!\right)\!-\!\left(\!\theta_2\!-\!\frac{f_{0,2}}{f_{0,1}\!+\!f_{0,2}}\!\right)\!\frac{\beta\!-\!\beta^k}{1\!-\!\beta}\\
0 & \theta_1\!\left(\!k-1\!\right)\!+\theta_2\frac{\beta-\beta^k}{1-\beta} & \theta_2\!\left(\!k-1\!\right)\!-\theta_2\frac{\beta-\beta^k}{1-\beta}\\
0 & \theta_1\!\left(\!k-1\!\right)\!-\theta_1\frac{\beta-\beta^k}{1-\beta} & \theta_2\!\left(\!k-1\!\right)\!+\theta_1\frac{\beta-\beta^k}{1-\beta}
\end{smallmatrix}
\end{pmatrix}.
\end{split}
\end{equation}

Solving (\ref{eq25}), we have
\begin{subequations}\label{eq27}
\begin{align}
\begin{split}
W_{n,0}\left(k\right)&=E\left[T_0\right]+\frac{1}{\hat{\mu}}\left(k-1\right)\\
&+\frac{\beta\!-\!\beta^k}{1\!-\!\beta}\left(E\left[T_0\right]-\frac{1}{\hat{\mu}}-\frac{1}{f_{0,1}+f_{0,2}}\right)
\end{split}\\
W_{n,1}\left(k\right)&=\frac{1}{\hat{\mu}}k+\frac{1\!-\!\beta^k}{1\!-\!\beta}\left(E\left[T_1\right]-\frac{1}{\hat{\mu}}\right)\\
W_{n,2}\left(k\right)&=\frac{1}{\hat{\mu}}k+\frac{1\!-\!\beta^k}{1\!-\!\beta}\left(E\left[T_2\right]-\frac{1}{\hat{\mu}}\right).
\end{align}
\end{subequations}

Let $W$ be the mean waiting time. According to (\ref{eq27}), we can derive the mean waiting time as follows:
\begin{align}\label{eq28}
\begin{split}
W =\!\sum_{j=0}^{2}\!\sum_{n=0}^{\infty}W\!_{n,j}\left(n\right)p_{n,j}\\
 =\frac{1}{1-\frac{\lambda}{\hat{\mu}}}\bigg[\frac{\lambda}{\hat{\mu}}E\left[T\right]+&\frac{1}{1\!-\!\beta}\sum_{j=0}^{2}E\left[T_j\right]\left(\pi_j-\hat{\pi}_j\right)\\
 &-\frac{\beta}{1\!-\!\beta}\frac{\tau}{r_C\tau\!+\!1}\left(\pi_0\!-\!\hat{\pi}_0\right)\bigg],
\end{split}
\end{align}
and thus the mean delay is
\begin{align}\label{eq29}
\begin{split}
D=W+E\left[T\right]\\
=\frac{1}{1-\frac{\lambda}{\hat{\mu}}}\bigg[E\left[T\right]&+\frac{1}{1\!-\!\beta}\sum_{j=0}^{2}E\left[T_j\right]\left(\pi_j-\hat{\pi}_j\right)\\
&-\frac{\beta}{1\!-\!\beta}\frac{\tau}{r_C\tau\!+\!1}\left(\pi_0\!-\!\hat{\pi}_0\right)\bigg].
\end{split}
\end{align}

Recall that the offloading efficiency, denoted by $\eta$, is defined as the ratio of the traffic transmitted by the Wi-Fi to the total traffic. In other words, the offloading efficiency is the proportion of the data that is served by the Wi-Fi on average in a frame. We define the time that a frame is served by the Wi-Fi as Wi-Fi service time and denote it by $U$. To solve $\eta$, we first derive the mean Wi-Fi service time, i.e., $E\!\left[U\right]$, using the hybrid embedded Markov chain developed.

Let $U_j$ be the Wi-Fi service time of a frame which starts its service in service state $j$. Consider a frame starting its service in the deferred state, i.e., state 0. In this case, the next embedded point will be $\Phi_j$ ($j=$1 or 2) with probability $f_{0,j}/\sum_{i=1}^{2}f_{0,i}$ , and the remaining Wi-Fi service time will be $U_j$. Therefore, we write the equation of $E\!\left[U_0\right]$ in (\ref{eq30a}):
\begin{subequations}\label{eq30}
\begin{align}
E\left[U_0\right]&=\frac{f_{0,1}}{f_{0,1}+f_{0,2}}E\left[U_1\right]+\frac{f_{0,2}}{f_{0,1}+f_{0,2}}E\left[U_2\right]\label{eq30a}.\\
\intertext{Similarly, we obtain the equations of $E\left[U_1\right]$ and $E\left[U_2\right]$:}
E\left[U_1\right]&=\frac{f_{1,2}}{\mu_1+f_{1,2}}E\left[U_2\right]\label{eq30b}\\
\begin{split}
E\left[U_2\right]&=\frac{\mu_2}{\mu_2+f_{2,0}}\frac{1}{\mu_2+f_{2,0}}\\
&+\frac{f_{2,0}}{\mu_2+f_{2,0}}\left(\frac{1}{\mu_2+f_{2,0}}+E\left[U_0\right]\right)\label{eq30c}.
\end{split}
\end{align}
\end{subequations}
Solving (\ref{eq30}), we have
\begin{subequations}\label{eq31}
\begin{align}
E\left[U_0\right]&=\frac{r_C\left(r_C\tau+\mu_1\tau+1\right)}{r_F\mu_1+(r_C\tau+1) r_C \mu_2+(r_C\tau+1) \mu_1 \mu_2}\\
E\left[U_1\right]&=\frac{r_C\left(r_C\tau+1\right)}{r_F\mu_1+(r_C\tau+1) r_C \mu_2+(r_C\tau+1) \mu_1 \mu_2}\\
E\left[U_2\right]&=\frac{\left(r_C+\mu_1\right)\left(r_C\tau+1\right)}{r_F\mu_1+(r_C\tau+1) r_C \mu_2+(r_C\tau+1) \mu_1 \mu_2}.
\end{align}
\end{subequations}
And thus the mean Wi-Fi service time is
\begin{equation}\label{eq32}
E\left[U\right]=\sum_{j=0}^{2}\hat{\pi}_jE\left[U_j\right].
\end{equation}

$E\left[U\right]$ times $\mu_2$ equals the average data amount served by the Wi-Fi in a frame (unit: frame, value range $[0,1]$), which is also the offloading efficiency
\begin{equation}\label{eq33}
\eta=\mu_2E\left[U\right].
\end{equation}
\subsection{Properties Observation}\label{4.2}
According to the theoretical results, we observe the effect of the deadline on the mean delay and the offloading efficiency. We also conduct simulation experiments of which the settings are the same as those in the M/MMSP/1 model. We consider the users in the vehicle as an example. We assume that the mean duration of channel state $C$ is $1/r_C=28.42$s and that of channel state $F$ is $1/r_F=12.57$s \cite{09}, which were obtained from the real trace data measured by \cite{17}. As the average frame size is 8.184kb, and the data rates of the cellular network and the Wi-Fi connections are respectively 8.7Mbps and 24.4Mbps \cite{14}, we set the frame arrival rate $\lambda=800$ frames/s, the frame service rates of the cellular network $\mu_1=8.7/0.008=1088$ frames/s and that of the Wi-Fi connections $\mu_2=24.4/0.008=3050$ frames/s.
\begin{figure}[htp]
\centering
\subfigure[Delay vs. deadline]{
 \label{fig8a}
 \includegraphics[scale=0.5]{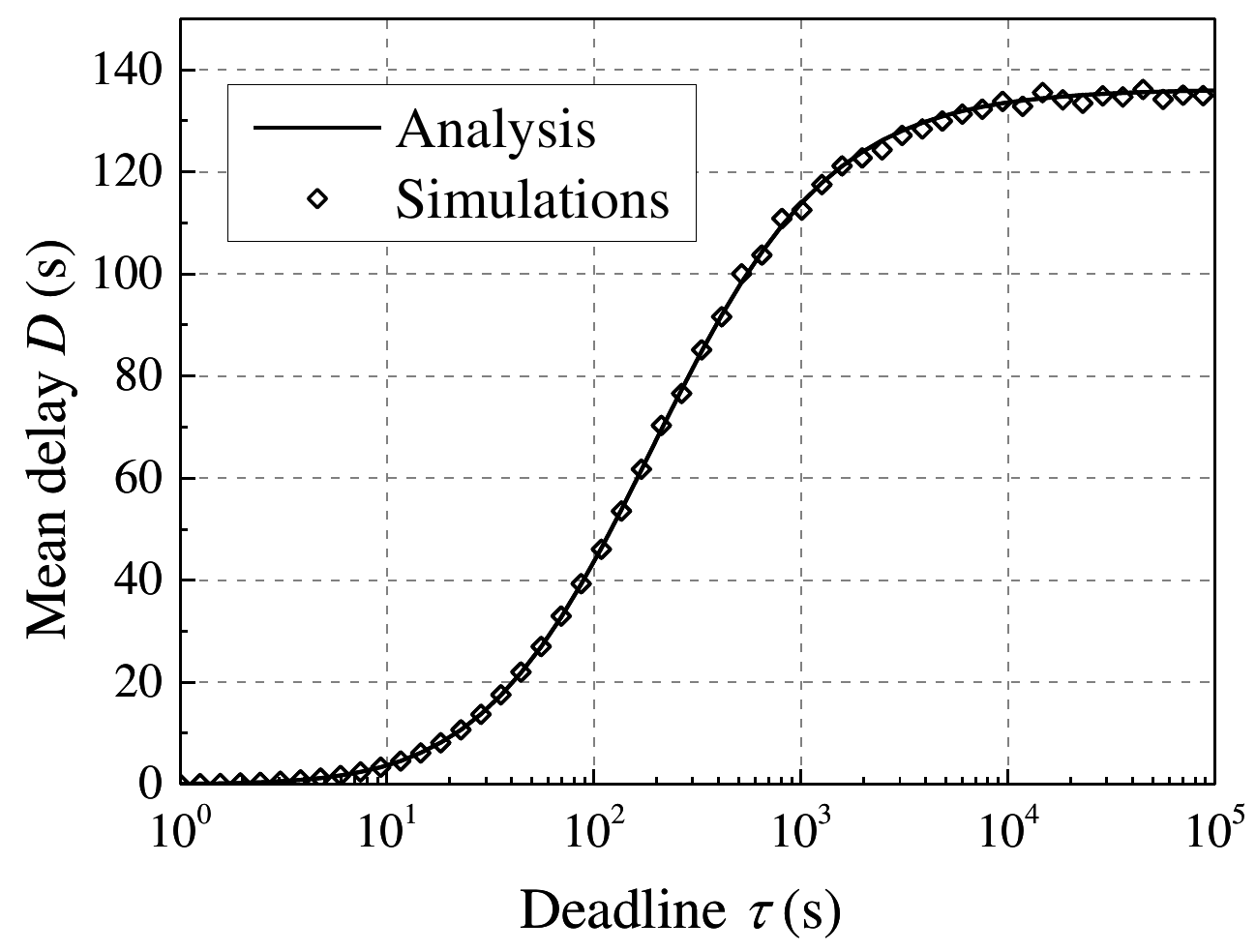}}
\subfigure[Efficiency vs. deadline]{{}
 \label{fig8b}
 \includegraphics[scale=0.5]{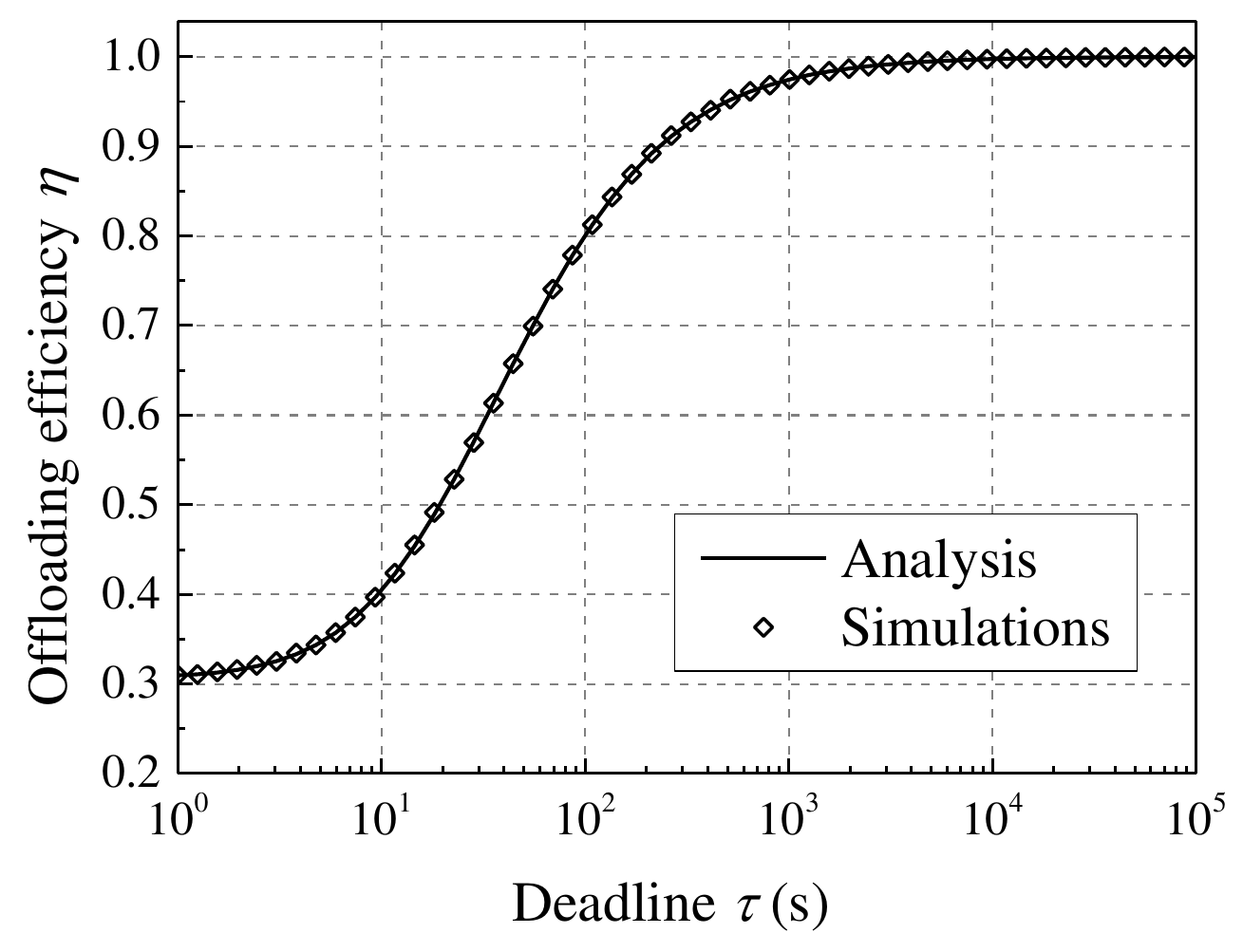}}
 \subfigure[Delay vs. efficiency]{
 \label{fig8c}
 \includegraphics[scale=0.5]{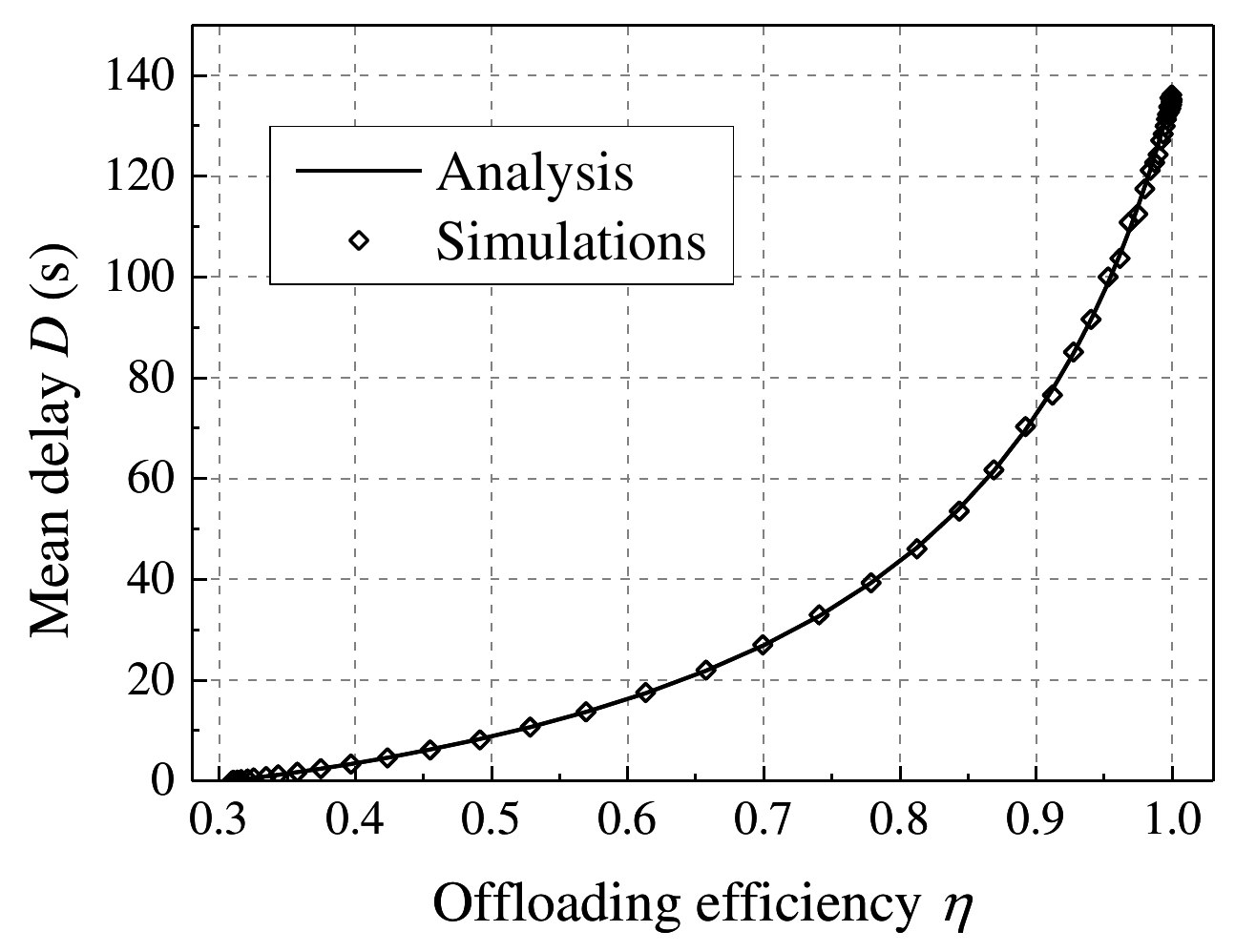}}
\caption{Delay and efficiency performance in the M/MMSP/1 queueing system.}
\label{fig8}
\end{figure}

We plot the mean delay $D$ and the offloading efficiency $\eta$ versus the deadline $\tau$ in Figure \ref{fig8a} and \ref{fig8b}. As we can see, both $D$ and $\eta$ monotonously increase with $\tau$, which means the offloading efficiency improved at the expense of the delay performance. In particular, when $\tau\rightarrow\infty$, $D$ converges to its maximum $\hat{D}$, which can be derived from (\ref{eq29}) as follows. 

When $\tau\rightarrow\infty$, the terminal transmits frames only when it has the Wi-Fi or all the traffic is transmitted via the Wi-Fi. In other words, there is no cellular state, i.e., $\lim\limits_{\tau\to\infty} \pi_1=0$ and $\lim\limits_{\tau\to\infty} \hat{\pi}_1=0$, and the service state transits only between the deferred state and the Wi-Fi state, of which the steady-state probabilities are $\lim\limits_{\tau\to\infty} \pi_0=1-R$ and $\lim\limits_{\tau\to\infty} \pi_2=R$, and the start-service probabilities $\lim\limits_{\tau\to\infty} \hat{\pi}_0=p_{0,0}$ and $\lim\limits_{\tau\to\infty} \hat{\pi}_2=1-p_{0,0}$. In this case, the system now reduces to an M/MMSP/1 queue with two service states in \cite{24}. Herein, the average service rate $\lim\limits_{\tau\to\infty} \hat{\mu}=(1-R)\cdot 0+R\cdot\mu_2=R\mu_2$, and the service is completely provided by the Wi-Fi state and thus $\lim\limits_{\tau\to\infty} \theta_2=1$. Consequently, We have the mean delay as follows
\begin{align}\label{eq34}
\hat{D}=&\lim\limits_{\tau\to\infty}D\nonumber\\
\begin{split}
=&\frac{1}{1\!-\!\lambda/\!\left(R\mu_2\right)\!}\Bigg\{p_{0,0}\frac{\mu_2\!+\!r_C\!+\!r_F}{r_C\mu_2}\\
&+\!\left(1-p_{0,0}\right)\!\frac{1}{R\mu_2}\!\!+\!\!\frac{\mu_1\!+\!r_C}{r_C}\!\bigg\{\frac{\mu_2\!+\!r_C\!+\!r_F}{r_C\mu_2}\!\Big[\!\left(1\!-\!R\right)\!-\!p_{0,0}\!\Big]\!\\
&+\frac{1}{R\mu_2}\Big[R\!-\!\left(1\!-\!p_{0,0}\right)\Big]\!\bigg\}\!-\!\frac{\mu_1}{r\!_C}\!\frac{1}{r\!_C}\!\left(1\!-\!R\!-\!p_{0,0}\right)\!\!\Bigg\}
\end{split}\nonumber\\
=&\frac{r_C+R\left(1-R\right)\mu_2}{r_C\left(R\mu_2-\lambda\right)}.
\end{align}
This indicates that $D$ approaches to a finite value $\hat{D}$ when $\tau\rightarrow\infty$ and $\lambda$ is less than $R\mu_2$, which is the capacity that the terminal can offer in this case. In the running example, the maximal mean delay $ \hat{D}=136.21$s, as Figure \ref{fig8a} shows. Moreover, we consider the offloading efficiency when the deadline approaches infinity. As all the data of a frame is served by the Wi-Fi service state with rate $\mu_2$, the conditional mean Wi-Fi service time
\begin{equation}\label{eq35}
\lim\limits_{\tau\rightarrow\infty} E\left[U_0\right]=\lim\limits_{\tau\rightarrow\infty} E\left[U_2\right]=\frac{1}{\mu_2}.
\end{equation}
It follows that the offloading efficiency
\begin{equation}\label{eq36}
\lim\limits_{\tau\rightarrow\infty}\eta=p_{0,0} \frac{1}{\mu_2}+0+\left(1-p_{0,0}\right)\frac{1}{\mu_2} =1,
\end{equation}
which is attributed to the fact that all the traffic is now transmitted via the Wi-Fi. 

Also, we plot the mean delay versus the offloading efficiency in Figure \ref{fig8c}. It is interesting to find that with the increase of offloading efficiency, the slope of mean delay monotonously grows. This indicates that though both $D$ and $\eta$ increase with $\tau$, the increasing speed of $D$ is larger than that of $\eta$. 

These properties imply that increasing $\tau$ from 0 to a small value can enhance the offloading efficiency with a small delay increment, which may improve the utility. However, if the terminal further increases $\tau$ to gain a large offloading efficiency, the mean delay will increase very quickly, which could instead lower down the utility. In other words, given the weight $a$, there exists an optimal deadline, denoted by $\tau^*$, to maximize the utility function $U$.

\begin{figure}[htp]
\centering
\subfigure[$a=0.9$]{
 \label{fig9a}
 \includegraphics[scale=0.5]{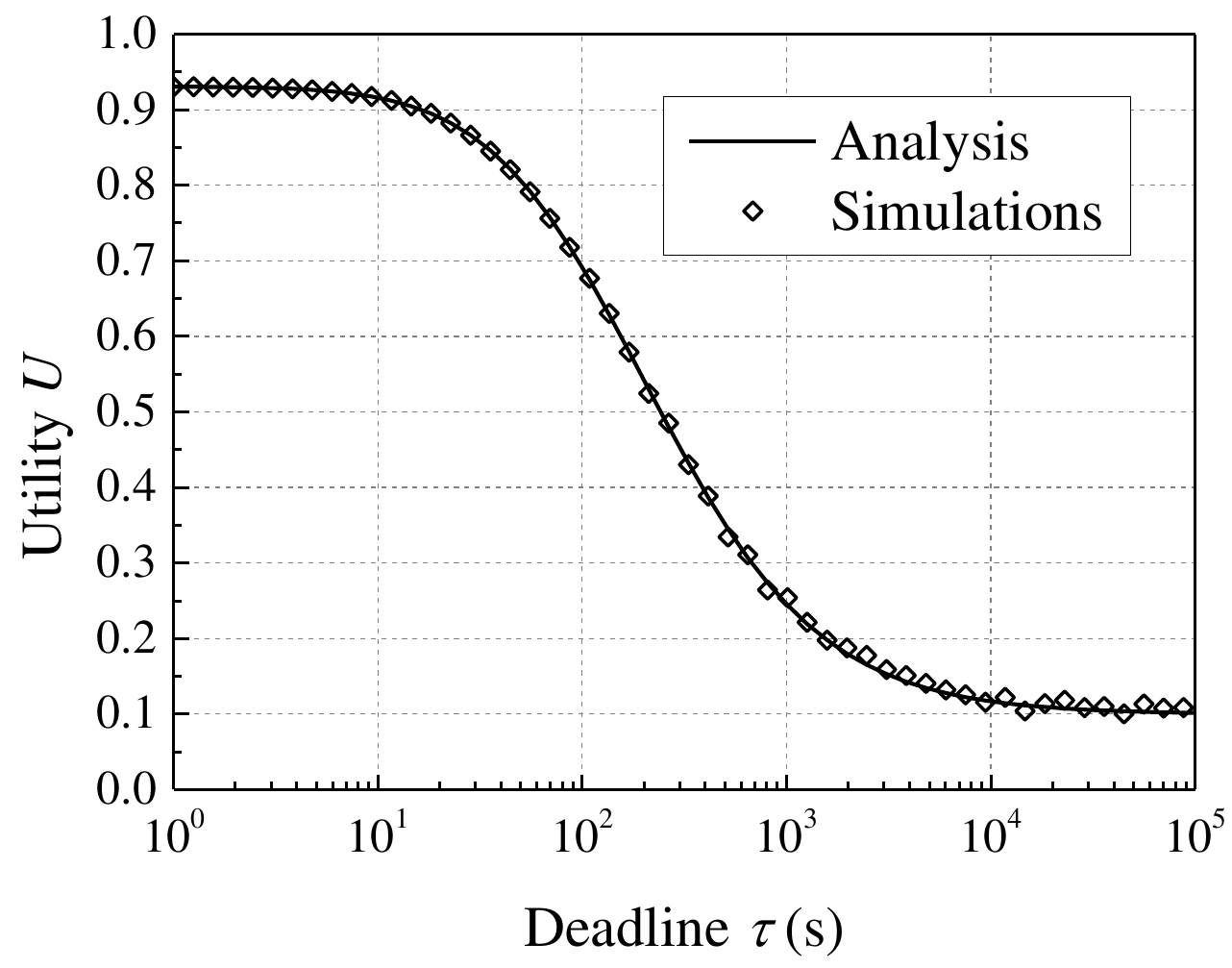}}
\subfigure[$a=0.5$]{{}
 \label{fig9b}
 \includegraphics[scale=0.5]{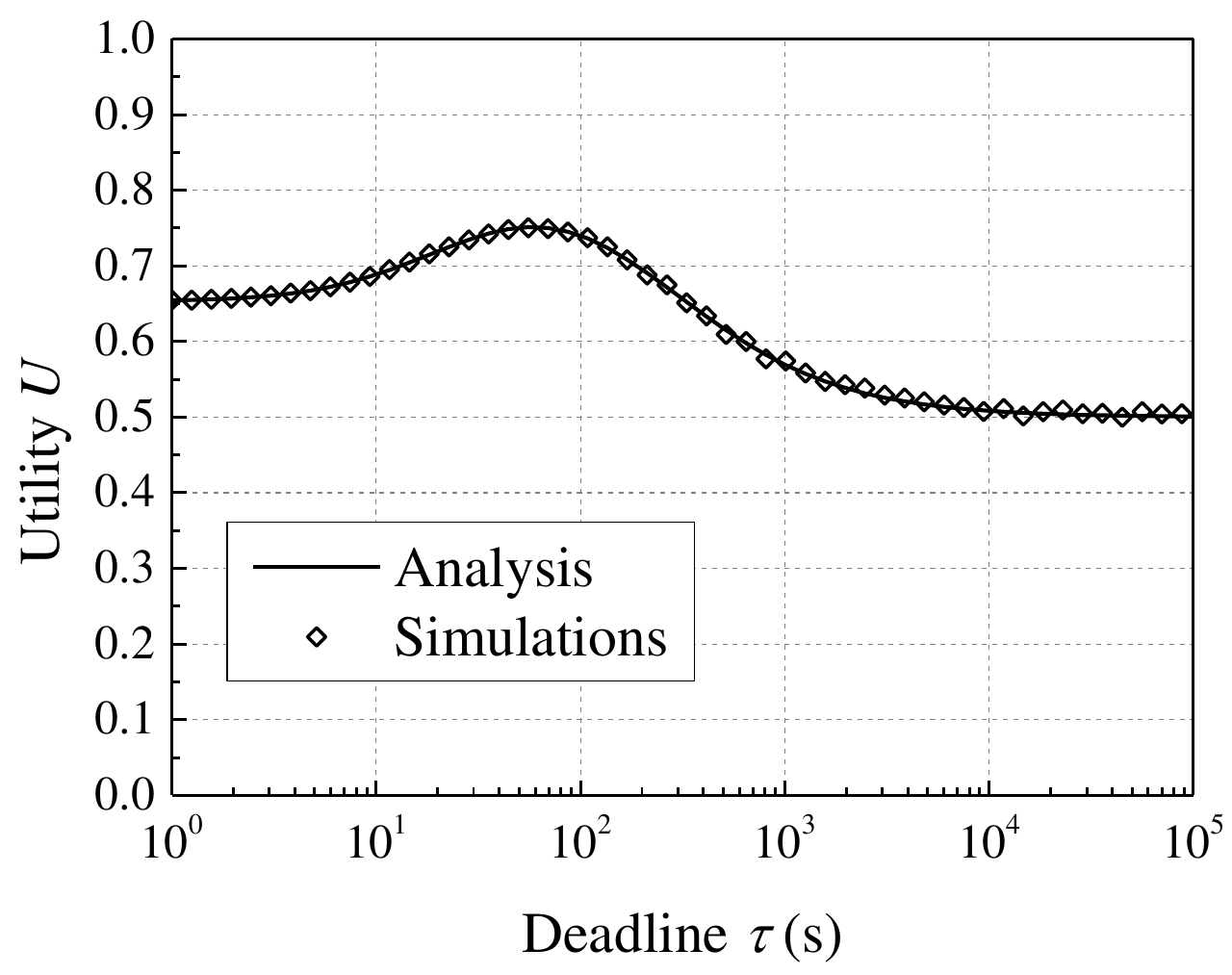}}
 \subfigure[$a=0.1$]{
 \label{fig9c}
 \includegraphics[scale=0.5]{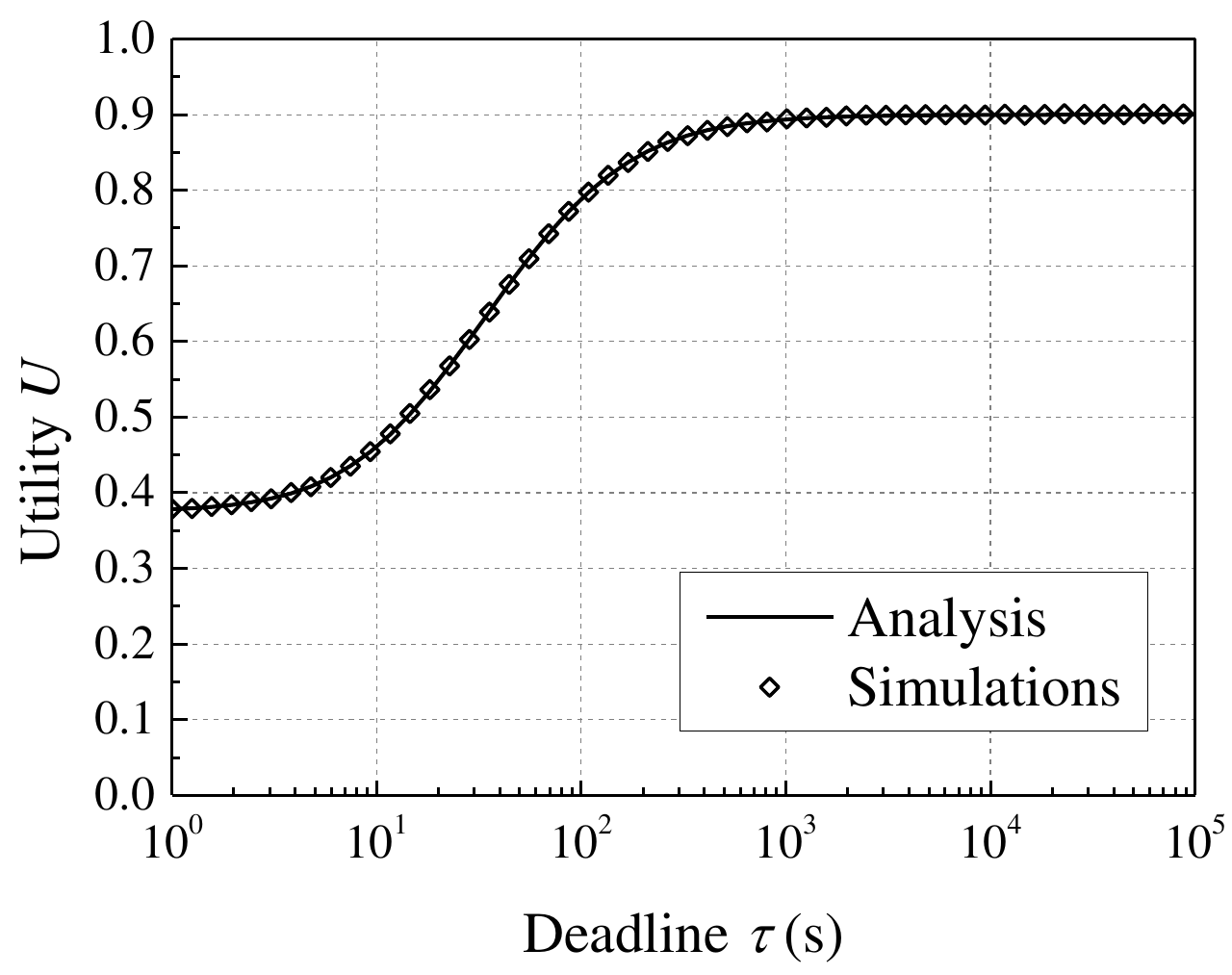}}
\caption{Utility vs. deadline in the M/MMSP/1 queueing system. }
\label{fig9}
\end{figure}
Substituting (\ref{eq29}), (\ref{eq33}), and (\ref{eq34}) into (\ref{eq1}), we can obtain the utility function in the M/MMSP/1 model. Based on this result, Figure \ref{fig9} studies the utility as a function of the deadline, where we consider the users with $a=0.9$, $a=0.5$, and $a=0.1$. When $a=0.9$, the user is highly sensitive to the delay performance and thus tends to select the on-the-spot offloading. As a result, the utility monotonically decreases with the deadline, and the optimal deadline $\tau^*\approx0$s. When $a=0.5$, the user is sensitive to the delay performance and the communication cost, and wants to make a tradeoff between them. In this case, the utility achieves its maximum at $\tau^*=55.5$s. When $a=0.1$, the user cares about the cost deeply and thus favors the pure offloading. In this case, the optimal deadline is $\tau^*=10^5$s.
\section{Applications}\label{applications}
According to the system properties, we provide the method to obtain the optimal deadline in Section \ref{5.1}, and examine the performance of the proposed strategy with the optimal deadline by simulations in Section \ref{5.2}.
\subsection{The method to obtain the optimal deadline}\label{5.1}
As we find in Section \ref{4.2}, there is only one peak of utility in the whole range of the deadline. Accordingly, we design a method to find out the optimal deadline. Given the average input traffic rate $\lambda$ and the preference weight $a$ of a user, the optimal deadline can be found according to the following procedure:
\begin{enumerate}[Step 1.]
\item Let $\tau=0$ and $U_a=U_b=0$.\label{step1}
\item Calculate $D$, $\eta$, $\hat{D}$, and $U_b$, using (\ref{eq29}), (\ref{eq33}), (\ref{eq34}), and (\ref{eq1}), respectively. If $U_b>U_a$, go to Step 3; otherwise, return $\tau$.\label{step2}
\item If $\tau<\hat{\tau}$, $\tau=\tau+\Delta\tau$, $U_a=U_b$, and jump to Step 2; otherwise return $\tau$.\label{step3}
\end{enumerate}
Herein, $\hat{\tau}$ is a large value, such as $10^5$ s, and $\Delta\tau$ is the increment step of $\tau$. Clearly, a small $\Delta\tau$ is helpful in finding an accurate optimal deadline at the expense of computational complexity.

In reality, our strategy can be implemented as a kind of cloud service. The wireless environment information of a city or region, such as the data rate of Wi-Fi hotspots and cellular network, as well as the transition rate of the duration times of two channels states, can be maintained in cloud servers in advance. Typically, such information is a kind of slowly varying information, which can be updated by cooperative users \cite{18}. When a user needs to update the optimal deadline, the user can send the preference weight $a$ and the long-run average traffic rate $\lambda$ to the cloud server. The server calculates the optimal deadline through the aforementioned method, and then sends it to the user. Since the changing frequency of the installation of Wi-Fi hotspots and cellular networks in a city or region are generally low, the update of the optimal deadline is not required to be very frequent. For example, to save the energy of the mobile terminals, the user can update the deadline once a month or when the user goes to a new city.
\subsection{Performance of the proposed strategy}\label{5.2}
The advantage of our proposal is that it can select an optimal deadline $\tau$ according to the preference of the user, such that a high utility can be achieved in the long run. To demonstrate this point, we verify the performance of our strategy by simulations, where compare four utility curves in total: 1) the utility under the on-the-spot offloading; 2) the utility under the pure offloading; 3) the utility under our strategy, where the optimal deadline is obtained from the method in Section \ref{5.1}; 4) the utility under our strategy, where the optimal deadline is obtained from simulations, which means we simulate the utility of our strategy with a wide variety of deadlines, and set the deadline with the maximal utility as the optimal deadline.

In the simulation, we change the preference weight $a$ from 0 to 1, to emulate the users with different sensitivities to the delay performance and the communication cost. The other parameters are the same as that in Figure \ref{fig8}.

\begin{figure}[htp]
\centering
\subfigure[] {
\label{fig10a}
 \includegraphics[scale=0.5]{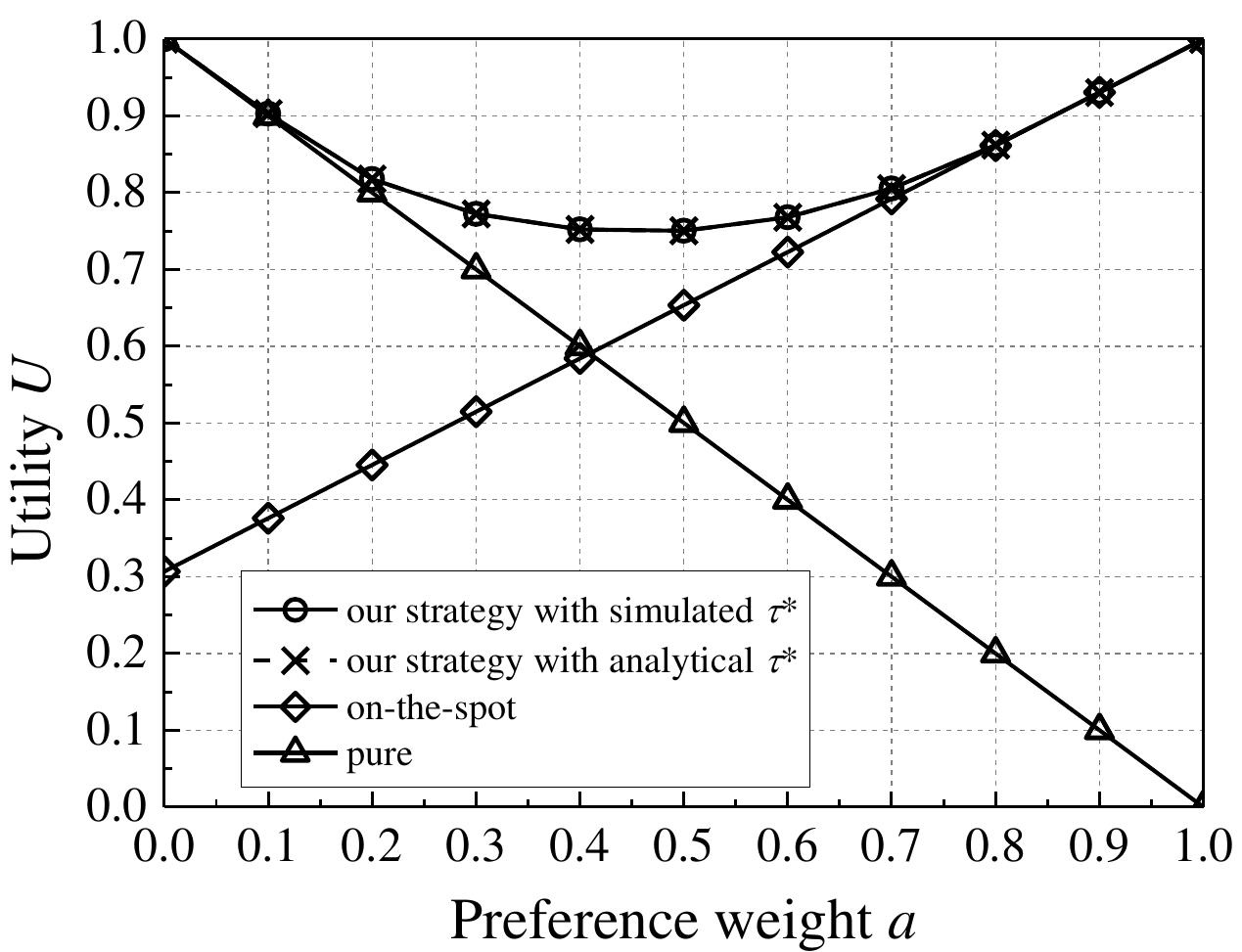}}
\subfigure[]{{}
 \label{fig10b}
 \includegraphics[scale=0.5]{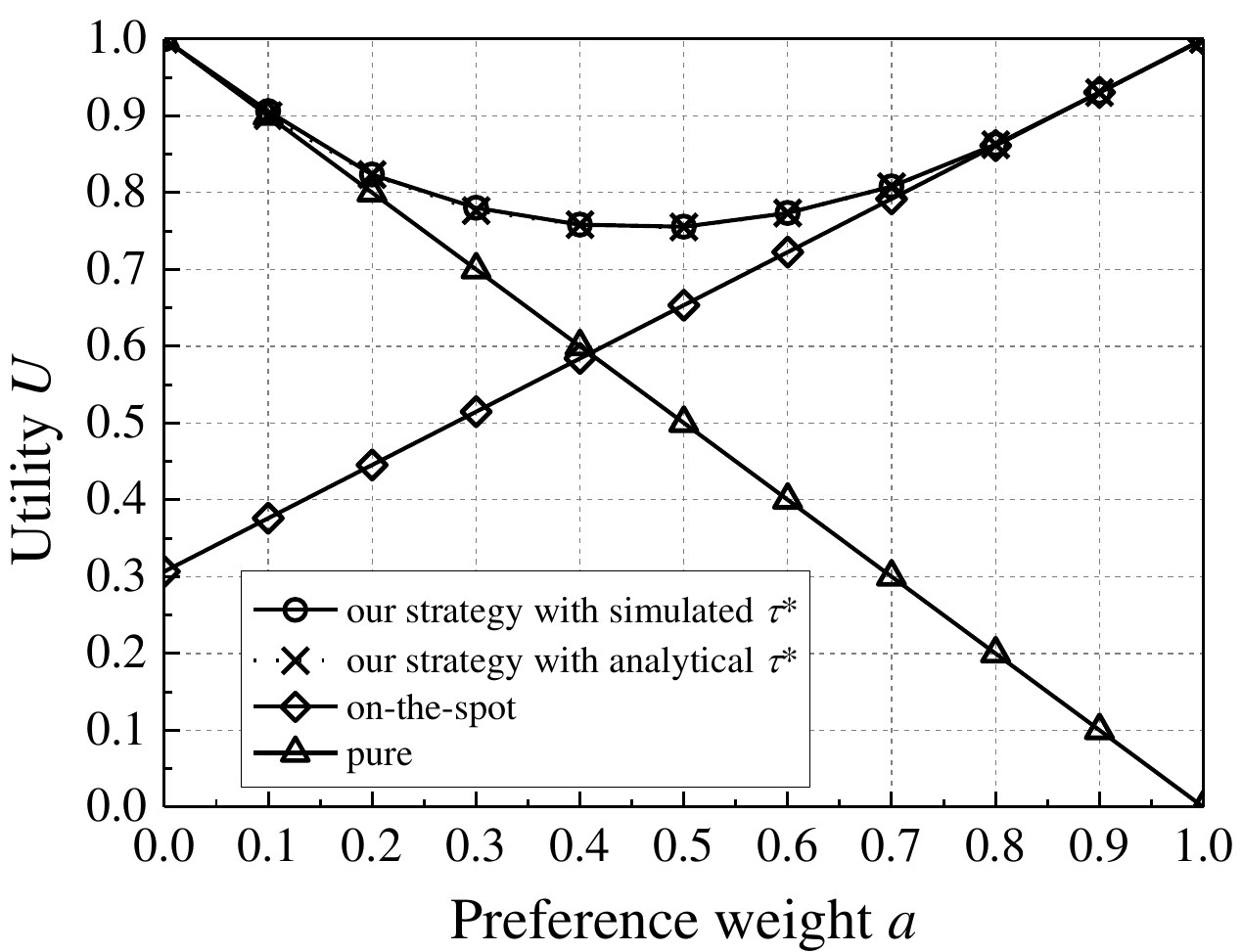}}
\caption{Utility $U$ vs. preference weight $a$ when the data rate of each Wi-Fi hotspot is (a) identical (b) different}
\label{fig10}
\end{figure}

Figure \ref{fig10a} provides the utility $U$ versus preference weight $a$ in simulation. As we can see, our strategy can always achieve a larger utility than the pure offloading and the on-the-spot offloading. When $a=0$, the user only cares about the communication cost and can bear a large mean delay. In this case, our strategy selects a large deadline, say $10^5$s, such that almost all the data can be transmitted via the Wi-Fi. As Figure \ref{fig10a} shows, our strategy can achieve a utility of 1 when $a=0$. With the increase of $a$, the user becomes more and more sensitive to the delay performance, and thus prefers to trade the communication cost for the delay performance. In this case, our strategy chooses a proper deadline according to the user's preference. For example, when $a=0.5$ in Figure \ref{fig10a}, our strategy chooses $\tau=35.53$s for the user and achieve a utility of 0.78. When $a=1$, the user cannot accept any interruption of data transmission. In this case, our strategy sets $\tau=0$s such that the terminal can switch to the cellular network immediately when it loses the Wi-Fi connection. As a result, it can obtain a utility of 1.

As a comparison, the pure offloading is a strategy with $\tau$ fixed at $\infty$, and the on-the-spot offloading is a strategy with $\tau$ fixed at 0s. As a result, the pure offloading and the on-the-spot offloading can achieve the same utility with our strategy only when $a=0$ and $a=1$, respectively. In other words, these two strategies cannot adapt to different users' preferences.

Furthermore, the speed of each Wi-Fi connection may be variable due to different router versions or the different load of the Wi-Fi hotspots, or the different distances between users and the Wi-Fi antennas in reality, etc. Thus, we further examine the performance of our strategy when the speed of each Wi-Fi hotspot is different. In the simulation, we set the speed of each Wi-Fi hotspot as a uniform random variable in the range [8.8 Mbps, 40 Mbps], of which the average is 24.4Mbps, but the speed of a single Wi-Fi hotspot is constant over time. As shown in Figure \ref{fig10b}, our strategy still achieves higher utility than that of two extreme offloading strategies.

Both Figure \ref{fig10a} and \ref{fig10b} show that, in our strategy, using the optimal deadline generated from theoretical results and using that selected from simulation bring similar utility. This phenomenon informs us that obtaining the optimal deadline from our theoretical results is feasible.
\section{Related works}\label{related}
Our proposed Wi-Fi offloading strategy aims at current commercial mobile terminals, which employ the Wi-Fi in preference to the cellular network when the Wi-Fi is available. Recently, there has emerged a new kind of terminals that can support the concurrent transmission mode \cite{16}. Such a kind of terminal transmit data via the Wi-Fi and the cellular network at the same time. An example is the download booster function of Samsung Galaxy S10 \cite{11}. Accordingly, several Wi-Fi offloading strategies \cite{06,07,08,09,10,17,20,21,12} have been proposed for such kinds of terminals. Similarly, the goal of these papers is to make a balance between the delay performance and the offloading efficiency.

In the strategy proposed by Cheng et al. \cite{06}, whether a data service will be offloaded via the Wi-Fi hotspot depends on the state of the Wi-Fi buffer when this data service is generated. If the Wi-Fi buffer is not full, the data will enter the Wi-Fi buffer and be eventually transmitted via the Wi-Fi; otherwise, it will be transmitted via the cellular network. The simulation results of \cite{06} demonstrated the tradeoff between the delay performance and the offloading efficiency.

In \cite{09}, \cite{08} and \cite{12}, the terminal performs data offloading through setting up a timer for each file when it is generated. The newly generated file first enters the Wi-Fi buffer. If the timer reaches a preset deadline and it is still in the Wi-Fi buffer, it will be transmitted via the cellular network. In particular, Mehmeti et al. \cite{09} provided the way to select an optimal deadline to minimize the mean delay or maximize the offloading efficiency when the traffic load is extremely large or small.

To improve the Quality of Experience (QoE), Ajith et al. \cite{10} added one prejudgment step to the Wi-Fi offloading strategy proposed by Lee et al. \cite{08}. When a new packet arrives, the system estimates how long it will wait before it becomes the HOL packet if it enters the Wi-Fi buffer. If the estimated waiting time is smaller than the preset timer, this packet will enter the Wi-Fi buffer; otherwise, it will be directly transmitted through the cellular network.

In \cite{17}, the terminal sets a delay-tolerant threshold for each application. Before transmitting the data of this application, the strategy predicts the volume of data that can be offloaded via the Wi-Fi before the delay threshold expires, according to historical information. If the predicted volume is not large enough, the data will be transmitted by the cellular network.

In \cite{07}, a handing-back point is set for the newly generated data, according to the history information of the moving paths of the user. If the terminal meets a Wi-Fi hotspot before the handing-back point, the data will be offloaded via the Wi-Fi hotspot; otherwise, it will be transmitted via the cellular network.

Zhang et al. proposed dynamic programming based Wi-Fi offloading strategies in \cite{20} and \cite{21}. In this kind of strategy, the time is slotted. In each time slot, the terminal decides whether the data should be delayed, transmitted via the cellular network, or offloaded via the Wi-Fi channel, according to the system state at the current slot, such as the amount of remaining data in the buffer and the location of the user.
\section{Conclusion}\label{conclusion}
This paper proposes a preference-oriented Wi-Fi data offloading strategy to achieve high utility in the long run for current commercial mobile terminals. In this strategy, the terminal pauses data transmission when it loses the Wi-Fi, and will resume data transmission via the cellular network if it cannot connect to a new Wi-Fi hotspot before the deadline expires. We develop a three-state M/MMSP/1 queueing model to depict this system and derive the structured expression of the mean delay and offloading efficiency by establishing a hybrid embedded Markov chain. Our analysis demonstrates that an optimal deadline can be selected according to the user's preference to maximize the utility. Our simulation results show that the proposed strategy can achieve larger utility than the on-the-spot offloading and the pure offloading.
\appendix 
\section{\texorpdfstring{Two-dimensional continuous-time Markov chain}{}} \label{appendixA}
\begin{figure}[htp]
\centering
\includegraphics[scale=0.86]{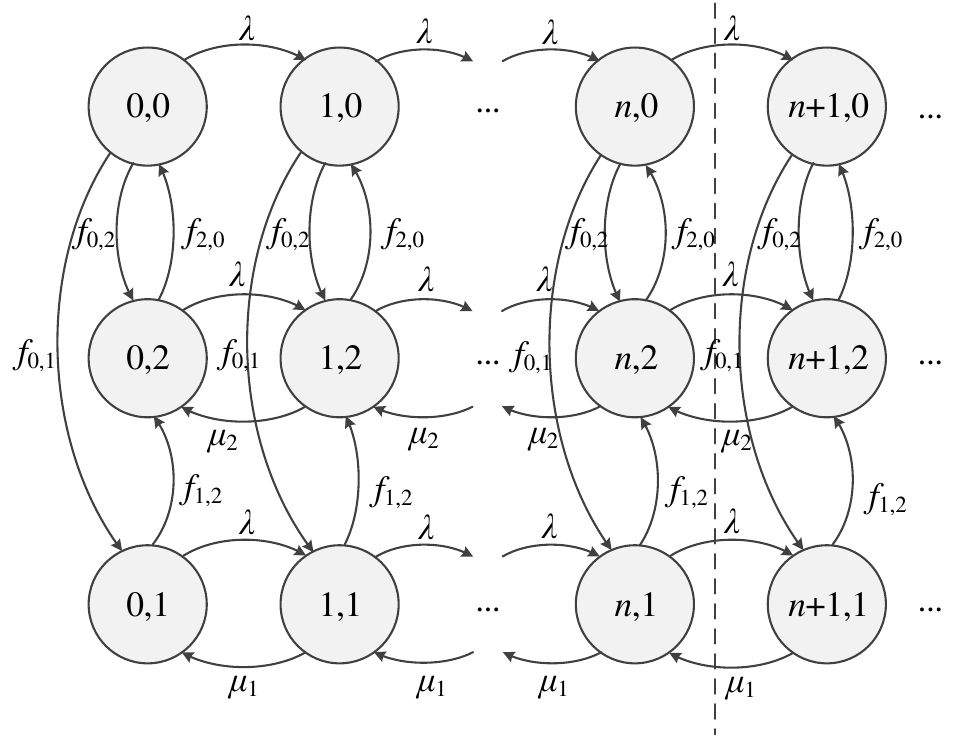}
\caption{State transition diagram of the two-dimensional Markov chain.}
\label{figa11}
\end{figure}

For an M/MMSP/1 queue, the system state can be determined by the service state and the number of data frames in the system. Let's define $X\left(t\right)$ as the number of data frames waiting in the buffer at time $t$ and $Y\left(t\right)$ as the service state at time $t$. The random process $\left\{\left(X\left(t\right),Y\left(t\right)\right),t\ge0\right\}$ is a two-dimensional continuous-time Markov chain with state space ${\left(n,j\right),n=0,1,2,\cdots,j=0,1,2}$. Figure \ref{figa11} plots the state transition diagram of the system.

After a long period of time, the system tends to be stable and the steady-state conditions is $\lambda < \hat{\mu}$ in (\ref{eq11}). We define the probability that $n$ frames in the buffer while the service state is $j$ as
\begin{equation}
p_{n,j}\!=\!\lim\limits_{t\rightarrow\infty}\!Pr\{X\!\left(t\right)\!=n,Y\!\left(t\right)\!=j;n\!=\!0,1,2,\!\cdots\!,j\!=\!0,1,2\}.
\end{equation}
It follows that the probability that there are $n$ frames in the buffer is
\begin{equation}
p_{n}=\lim\limits_{t\rightarrow\infty}Pr\{X(t)=n\}=\sum_{j=0}^{2}p_{n,j}.
\end{equation}
According to the service state, we define the partial generating functions as
\begin{equation}
G_j (z)=\sum_{n=0}^{\infty}p_{n,j} z^n,\left|z\right|\le1,j=0,1,2,
\end{equation}
and the overall generation function of the queue length is
\begin{equation}
G\left(z\right)=\sum_{n=0}^\infty p_n z^n=\sum_{j=0}^{2}G_j\left(z\right).
\end{equation}
According to the state transition diagram in Figure \ref{figa11}, the equilibrium equations of this continuous-time Markov chain can be given as follows:
\begin{subequations}\label{eqa1}
\begin{align}
&\left(\lambda+f_{0,1}+f_{0,2}\right)p_{0,0}=f_{2,0}p_{0,2}\\
&\left(\lambda+f_{2,0}\right)p_{0,2}=\mu_2p_{1,2}+f_{0,2} p_{0,0}+f_{1,2} p_{0,1}\\
&\left(\lambda+f_{1,2}\right) p_{0,1}=\mu_1p_{1,1}+f_{0,1} p_{0,0},  
\end{align}
\end{subequations}
and
\begin{subequations}\label{eqa2}
\begin{align}
&\left(\lambda+f_{0,1}+f_{0,2}\right)p_{n,0}=\lambda p_{n-1,0}+f_{2,0}p_{n,2}\\
&\!\left(\lambda\!+\!f_{2,0}\!+\!\mu_2\right)\!p_{n,2}\!=\!\lambda p_{n-1,2}\!+\!\mu_2p_{n+1,2}\!+\!f_{0,2} p_{n,0}\!+\!f_{1,2} p_{n,1}\\
&\left(\lambda+f_{1,2}+\mu_1\right) p_{n,1}=\lambda p_{n-1,1}+\mu_1p_{n+1,1}+f_{0,1} p_{n,0},  
\end{align}
\end{subequations}
where $n\ge1$. From (\ref{eqa1}) and (\ref{eqa2}), $G_j\!\left(z\right)$ can be obtained as follows
\begin{subequations}\label{eqa3}
\begin{gather}
\begin{split}\label{eqa3a}
G_0\left(z\right)=&r_F\tau\Big\{-\mu_2 p_{0,2}\lambda z^2+\big[\left(\mu_1+\lambda\right)\mu_2 p_{0,2}\\
&+r_C \left(\hat{\mu}-\lambda\right)\big]z-\mu_1 \mu_2 p_{0,2}\Big\}/g\left(z\right),
\end{split}\\
\intertext{and}
G_1\left(z\right)=\frac{1}{\tau} \frac{(\hat{\mu}-\lambda-\mu_2 p_{0,2} )\tau\left(z-1\right)+zG_0\left(z\right)}{-\lambda z^2+(\lambda+r_C+\mu_1 )z-\mu_1}\label{eqa3b}\\
G_2\left(z\right)=\frac{-\lambda\tau z+(r_C+\lambda)\tau+1}{r_F\tau} G_0 \left(z\right)\label{eqa3c},
\end{gather}
\end{subequations}
where
\begin{equation}\label{eqa4}
\begin{split}
g(z)\!\!=&\!\Big\{\!\!\!-\!\lambda\tau\!\left[\!-\!\lambda z^2\!+\!\left(\lambda\!+\!r_F\!+\!\mu_2\right)z\!-\!\mu_2\right]\!+\!\left(\!r_C \tau\!+\!1\!\right)\!\left(\!\mu_2-\lambda z\!\right)\!\!\Big\}\!\\
&[\!-\!\lambda z^2\!+\!(\lambda\!+\!r_C\!+\!\mu_1 )z\!-\!\mu_1 \big]\!+\!r_F \!\left(\mu_1\!-\!\lambda z\right)\!z.
\end{split}
\end{equation}

A unknown parameter $p_{0,2}$ is left in the numerators of (\ref{eqa3}). If we put the roots of the denominator of (\ref{eqa3a}) into the numerator, the numerator should be zero, otherwise $G_0\left(z\right)$ will approach to infinity and the system will be unstable. We substitute $z=0$ and $z=1$ in $g\left(z\right)$, and get $g\left(0\right)=-\mu_1\mu_2 \left(\lambda+r_C+1/\tau\right)<0$ and $g\left(1\right)=\left(r_C+r_F\right)\left(r_C\tau+1\right)\left(\hat{\mu}-\lambda\right)/\tau>0$. It follows that there must exists one or more root(s) of $g\left(z\right)$ between 0 and 1, which is/are valid root(s) satisfing $\left|z\right|\le1$. Denote one of the valid roots by $z_0$, which can be solved numerically. Put it into the numerator of $G_0\left(z\right)$, we have 
\begin{equation}\label{eqa5}
p_{0,2}=\frac{r_C\left(\hat{\mu}-\lambda\right)z_0}{\mu_2\lambda z_0^2-\left(\mu_1+\lambda\right)\mu_2z_0+\mu_1\mu_2}.
\end{equation}
Substitute $p_{0,2}$ in (\ref{eqa3}), we get the expression of $G_j\left(z\right)$ respect to $z$. By letting $z=0$ in $G_0\left(z\right)$, we can get the steady-state probability that the service state is deferred state and the system is empty
\begin{equation}\label{eqa6}
p_{0,0}=G_0\left(0\right).
\end{equation}
The generating function $G(z)=\sum_{j=0}^{2}G_j\left(z\right)$. According to Little's law, we have the mean delay
\begin{equation}\label{eqa7}
\begin{split}
&D=\frac{G'\left(1\right)}{\lambda}=\frac{1}{d\left(z\right)}\bigg\{\!-\!\lambda\!\left(\!r_C\!+\!r_F\!\right)\!^2\!-\!\left(\!\mu_2\!-\!\mu_1\right)\! r_C \!\left(\!\mu_2-\lambda\right)\!\\
&+r_C\Big\{-2\lambda\!\left(\!r_C\!+\!r_F\right)^2\!+\!\left(\!\mu_2\!-\!\lambda\!\right)\!\big[\mu_1 r_F\!-\!2r_C \!\left(\!\mu_2\!-\!\mu_1 \!\right)\!\big]\Big\}\tau\\
&-r_C \Big\{\lambda\!\left(\!r_C\!+\!\mu_1\!\right)\!\left(\!r_F\!+\!r_C\!\right)^2\!+\!\mu_2 r_C\!\big[r_C \!\left(\!\mu_2\!-\!\lambda\!\right)\!-\!\mu_1 \!\left(\!r_C\!+\!r_F \!\right)\!\big]\Big\}\tau^2\\
&-\!\mu_2\!\left (\!r_C \tau \!+\!1\!\right)\!\left(\!r_C\!+\!r_F \right)\!\big[\mu_1 r_F \tau\!-\!\left(\!\mu_2\!-\!\mu_1\! \right)\!\left(\!r_C \tau\!+\!1\!\right)\!\big] p_{0,2} \!\bigg\},
\end{split}
\end{equation}
where
\begin{equation}\label{eqa8}
\begin{split}
d\left(z\right)=&\lambda\!\left(r_C\!+\!r_F\right)\!\left(r_C\tau\!+\!1\right)\!\big[\lambda\left(r_C\!+\!r_F\right)\!\left(1\!+\!r_C\tau\right)\\
&-r_C\mu_2\!\left(1\!+\!r_C\tau\right)\!-\mu_1 r_F\big]\!.
\end{split}
\end{equation}

This expression is a complicated combination of mathematical notations without any intermediate parameters, thus gives little physical insights in studying the proposed strategy.

\bibliographystyle{ieeetr}
\bibliography{mybibfile}

\end{document}